\documentclass[twocolumn]{aastex701}

\usepackage{amsmath}
\begin{document}

\title{Extreme-ultraviolet synthesis of nanojet-like ejections due to coalescing flux ropes}
\shorttitle{EUV synthesis of nanojet-like ejections}

\author[orcid=0000-0003-1546-381X, sname='Sen']{Samrat Sen$^*${\includegraphics[scale=0.07]{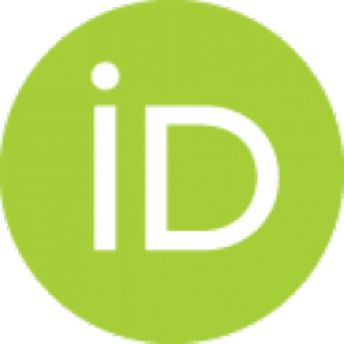}}\href{https://orcid.org/0000-0003-1546-381X}}
\affiliation{Instituto de Astrof\'{i}sica de Canarias, 38205 La Laguna, Tenerife, Spain}
\affiliation{Universidad de La Laguna, 38206 La Laguna, Tenerife, Spain}
\email[show]{$^*$samratseniitmadras@gmail.com; $^*$samrat.sen@iac.es} 
\author[orcid=0000-0002-9723-3155]{A. Ramada C. Sukarmadji{\includegraphics[scale=0.07]{figures/orcid-ID.pdf}}}
\affiliation{Institut de Recherche en Astrophysique et Planétologie, 9, avenue du Colonel Roche, 31028, Toulouse, France}
\email{asukarmadji@irap.omp.eu} 
\author[orcid=0000-0002-7788-6482]{D. N{\'o}brega-Siverio{\includegraphics[scale=0.07]{figures/orcid-ID.pdf}}}
\affiliation{Instituto de Astrof\'{i}sica de Canarias, 38205 La Laguna, Tenerife, Spain}
\affiliation{Universidad de La Laguna, 38206 La Laguna, Tenerife, Spain}
\email{dnobrega@iac.es} 
\author{F. Moreno-Insertis}
\affiliation{Instituto de Astrof\'{i}sica de Canarias, 38205 La Laguna, Tenerife, Spain}
\affiliation{Universidad de La Laguna, 38206 La Laguna, Tenerife, Spain}
\email{fmi@iac.es} 
\author[orcid=0000-0002-0333-5717]{J. Mart{\'i}nez-Sykora{\includegraphics[scale=0.07]{figures/orcid-ID.pdf}}}
\affiliation{SETI Institute, 339 Bernardo Ave, Mountain View, CA 94043, USA}
\affiliation{Lockheed Martin Solar and Astrophysics Laboratory, 3251 Hanover Street, Palo Alto, CA 94306, USA}
\affiliation{Institute of Theoretical Astrophysics, University of Oslo, P.O. Box 1029 Blindern, N-0315 Oslo, Norway}
\affiliation{Rosseland Centre for Solar Physics, University of Oslo, P.O. Box 1029 Blindern, N-0315 Oslo, Norway}
\email{jmsykora@seti.org} 
\author[orcid=0000-0003-1529-4681]{Patrick Antolin{\includegraphics[scale=0.07]{figures/orcid-ID.pdf}}}
\affiliation{School of Engineering, Physics, and Mathematics, Northumbria University, Newcastle upon Tyne, NE1 8ST, UK}
\email{patrick.antolin@northumbria.ac.uk} 
\shortauthors{Sen et al.}

\begin{abstract}
Detection and characterization of small-scale energetic events such as nanoflares and nanojets remain challenging owing to their short lifetimes, small spatial extent, and relatively low energy release, despite their potential role in coronal heating. Recent observations have identified nanojets as small-scale (length $\lesssim 6.6$~Mm, width $\lesssim 1$~Mm), fast ($\sim$~few 100 km s$^{-1}$), and short-lived ($\lesssim 30$~s) ejections associated with nanoflare-scale energies, providing evidence of magnetic reconnection at small spatial scales. However, the lack of synthetic diagnostics has limited the connection between magnetohydrodynamic (MHD) models and observations. In this Letter, we present synthetic observations of the coalescence of two flux ropes, leading to nanojet-like signatures from a numerical model obtained with the \texttt{MPI-AMRVAC} code. We report synthetic observables in Extreme-ultraviolet lines compatible with existing instruments such as SDO/AIA, and upcoming MUSE mission, and compare the synthetic observables with an existing observation of nanojets. The synthetic diagnostics of the emissivity maps, Doppler velocity, thermal, and non-thermal line broadening produce key observational properties, suggesting a plausible 3D scenario for nanojet generation where tiny flux ropes reconnect within loops. Our results provide predictions for the detectability of nanojets with current and future spectroscopic facilities, and establish a bridge between MHD modeling and observations.

\end{abstract}

\keywords{line: formation; magnetic reconnection; magnetohydrodynamics (MHD); methods: numerical; Sun: corona}

\section{Introduction} \label{sec:intro}
Detection of small-scale energetic events such as nanoflares and nanojets is challenging with the current state-of-the-art observing facilities, as they demand high spatial ($\lesssim$~few hundreds of km) and temporal ($\lesssim$~tens of seconds) resolution to capture them. Despite their short lifetimes, sizes, and low energy release, these small-scale events (nanoflares) are expected to occur at a rate of $10^3-10^4$~s$^{-1}$ \citep{Pauluhn:2007}, and are believed to significantly contribute to heating and maintaining the solar corona to multi-million-kelvin temperatures. 

The discovery of nanojets by \cite{Antolin:2021nanojet} revealed the observational signature of magnetic reconnection due to field line braiding, which is also supported by the observations by \cite{Ramada:2022} and \cite{Ramada:2024}. They report that these are small-scale ($\lesssim 1.5$~Mm in length, $\lesssim 1$~Mm in width), short-lived ($\lesssim 20$~s), and fast ($\approx~100$~km s$^{-1}$) ejections which correspond to the nanoflare energy range of $10^{22}-10^{26}$~erg \citep[][and references therein]{Fludra:2023, Belov:2024}. On the other hand, using high-resolution observations in the extreme-ultraviolet (EUV) range with Solar Orbiter (SolO) / HRI$_\mathrm{EUV}$ \citep{SolO-EUI:2020}, statistical surveys of nanojets prior to and during prominence eruption were reported by \cite{Gao:2025}, \citet{Bura_2025ApJ...988L..65B}, \citet{Tan_2025arXiv250904741T} and \citet{Wallace:2025}. These studies found two classes of nanojets. The first class matches the IRIS \citep[\textit{Interface Region Imaging Spectrograph};][]{IRIS:2014} observations reported by \cite{Antolin:2021nanojet, Ramada:2022}. The second class is much more energetic (high speeds up to $1000$~km s$^{-1}$), longer and numerous, suggesting that nanojet properties may scale with the total energy release.

Another puzzling feature of nanojets is that the large majority of them have been found to be unidirectional, in apparent contradiction with the interpretation of small-angle reconnection. \citet{Antolin:2021nanojet} conjectured that this could be due to the local curvature of the loop or, more generally, to the lack of symmetry around the current sheet. Indeed, \cite{Antolin:2021nanojet} and \cite{Ramada:2022} reported that most of the nanojets propagate inward relative to the curvature radius of their hosting coronal loops. In contrast, \cite{Ramada:2024} reported the observational evidence of outward-directed nanojets in a coronal loop, and \cite{Gao:2025} provided evidence of nanojet ejections occurring in two opposite directions with nearly equal occurrence during a prominence eruption, where the effect of field line curvature is minimal, resulting in no strong directional bias. However, nanojets have been found to be directed even perpendicular to the loop plane \citep{Ramada:2022}, suggesting that the loop's internal braiding may be the key to the puzzle. Nevertheless, there are still many discrepancies between our theoretical understanding and the observations in the context of nanojet generation mechanism and their properties, such as the directional bias of the ejections, the reason for single and clustered occurrences, the associated plasmoid-like ejecta, the narrow widths, and the high speeds up to $1000$ km s$^{-1}$. 

An analytical model by \cite{Pagano:2021} showed that ejections of the nanojet type can be asymmetric when curvature is included, with a more energetic flow along the curvature radius, due to reconnection between two slightly misaligned coronal loops. They also supported these phenomena using 2.5D MHD simulation. In contrast, the 3D MHD simulation by \cite{Antolin:2021nanojet} demonstrated that such ejections end up being mostly bi-directional, arising from the coalescence of two straight misaligned flux tubes. The modeling effort in 3D by \cite{Daughton:2011} demonstrated the formation of flux ropes through the development of the tearing instability in a current sheet (CS), where they reported the development of secondary flux ropes and their subsequent interactions. A 2D model by \cite{2022A&A...666A..28S}, and 3D models by \cite{2023A&A...678A.132S, DeJonghe:2025} incorporated the non-adiabatic effects of thermal conduction, radiative cooling, and background heat to realistically model the solar corona. They demonstrated the coupled tearing–thermal evolution in coronal CSs, capturing the formation of flux ropes (plasmoids in 2D), and the multi-thermal (from $\sim$~kK to MK) corona. However, none of these models reported jet-like ejections resulting from the coalescence of plasmoids or flux ropes. 

More recently, \citet[][hereafter SMI25]{Sen-FMI:2025} demonstrated, for the first time, that bi-directional nanojet-like ejections can occur due to reconnection at the interface of two merging flux ropes. However, the lack of synthetic observations of these ejections due to flux rope coalescence, and a comparative study between the synthetic observations and existing nanojet observations has not received attention to date. Therefore, it is of interest to produce synthetic observations compatible with existing and future telescopes to compare the synthetic observables with existing observations and to guide future observational campaigns of nanojets. 

To that end, in this work, we produce synthetic images and spectral line profiles focusing on the flux rope merging and nanojet-like ejections in different EUV lines compatible with the existing \textit{Atmospheric Imaging Assembly} (AIA) instrument onboard Solar Dynamics Observatory \citep[SDO;][]{Lemen_AIA:2012} and the upcoming Multi-Slit Solar Explorer \citep[MUSE;][]{MUSE-Depontieu:2020} from the MHD model by SMI25. Furthermore, we compare the morphological similarities between the synthetic images with the observation by \cite{Ramada:2022}, which showed the interaction of two plasmoid-like structures prior to the nanojet ejection for a blowout jet scenario in different EUV channels of SDO/AIA. 

\section{Summary of the underlying MHD model}\label{sec:simulation}

In the numerical simulation by \cite{Sen-FMI:2025} that serves as a basis for the synthesis carried out in this letter, we investigate small-scale ejections arising from the coalescence of magnetic flux rope pairs in a coronal current sheet (CS) using a 2.5D resistive-MHD simulation with \texttt{MPI-AMRVAC} \citep{2012JCoPh.231..718K, 2014ApJS..214....4P, 2018ApJS..234...30X, keppens2021, keppens2023}. The system is initialized in a force-free state within a stratified, low-$\beta$ ($\lesssim 0.3$) corona under solar gravity, that includes field-aligned thermal conduction. The computational domain extends from $x=-20$ to 20 Mm, $y=0$ to 40 Mm, and is invariant in $z$. Initially, $B_x=0$ throughout the domain, $B_y$ reverses its sign across $x=0$ (at the CS location), where the guide field peaks at $B_z=6$ G. Implementation of adaptive mesh refinement yields the highest resolution of 26.04 km along either direction. The model comprises a uniform diffusivity of $2.33 \times 10^{12}$~cm$^2$ s$^{-1}$ and the associated Lundquist number of $\approx 1.3 \times 10^4$, which is within the regime of the plasmoid instability \citep{2007PhPl...14j0703L}. However, the magnetic diffusivity in our model is chosen much larger than the typical coronal value of $\sim 10^4$~cm$^2$ s$^{-1}$ \citep{Vekstein:2016}, which is a standard practice of coronal modeling in resistive-MHD regime \citep[][and references therein]{Ni:2012, Shen:2022NatAs, Sen:2024, Sen:2025b}. Use of the actual coronal diffusivity in the MHD models would demand spatial resolutions of the simulation many orders of magnitude higher than used in our model, which is extremely challenging (if not impossible) to implement. The readers are referred to SMI25 for a further description of the initial condition, numerical schemes, and boundary conditions used in the model.

\begin{figure*}[hbt] 
    \centering
    \includegraphics[width=0.8\linewidth]{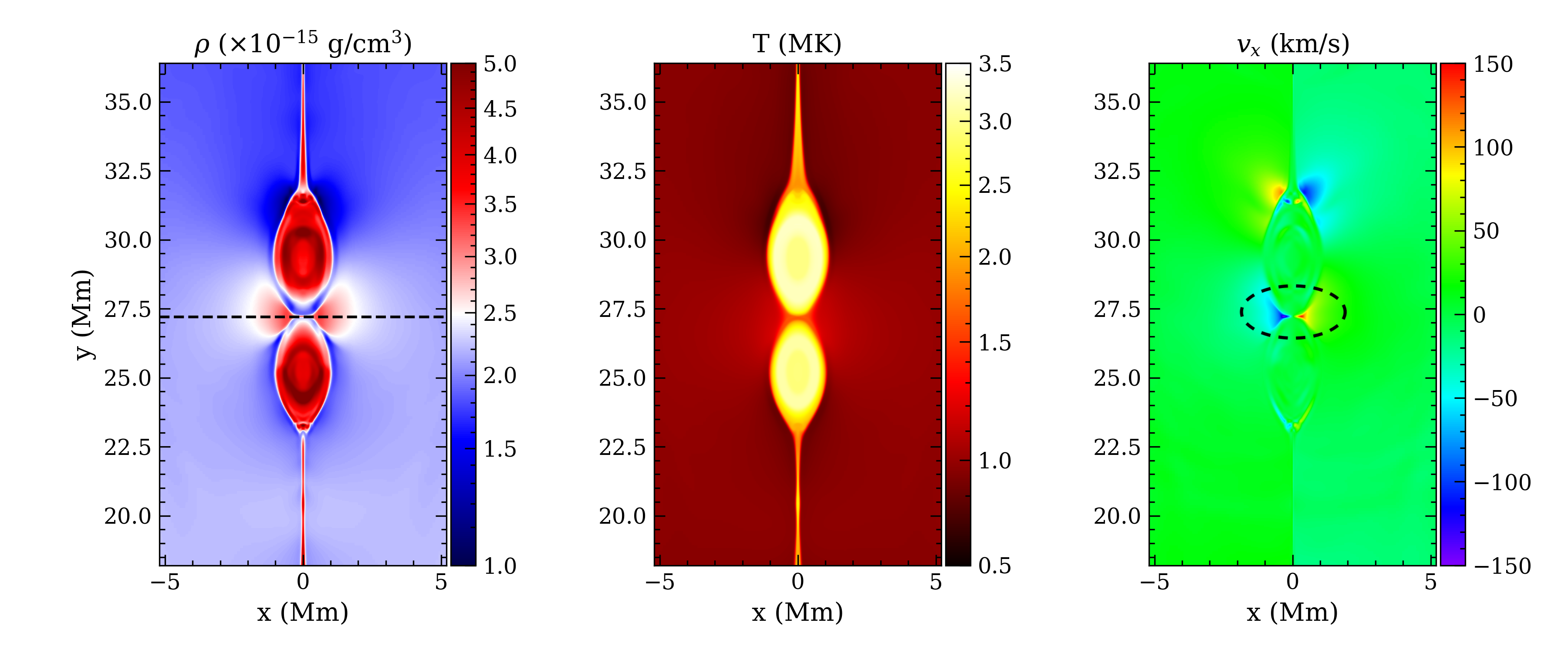}
    \includegraphics[width=0.8\linewidth]{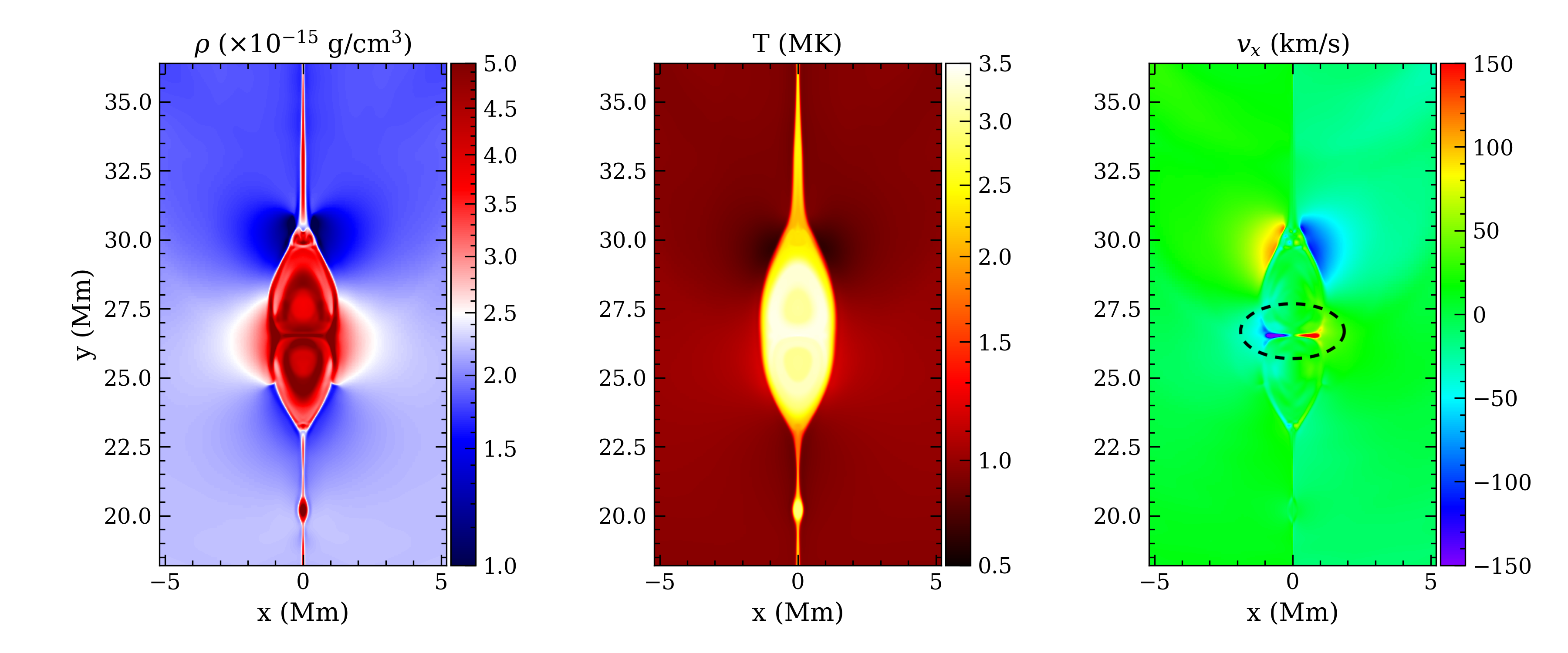}
    \includegraphics[width=0.8\linewidth]{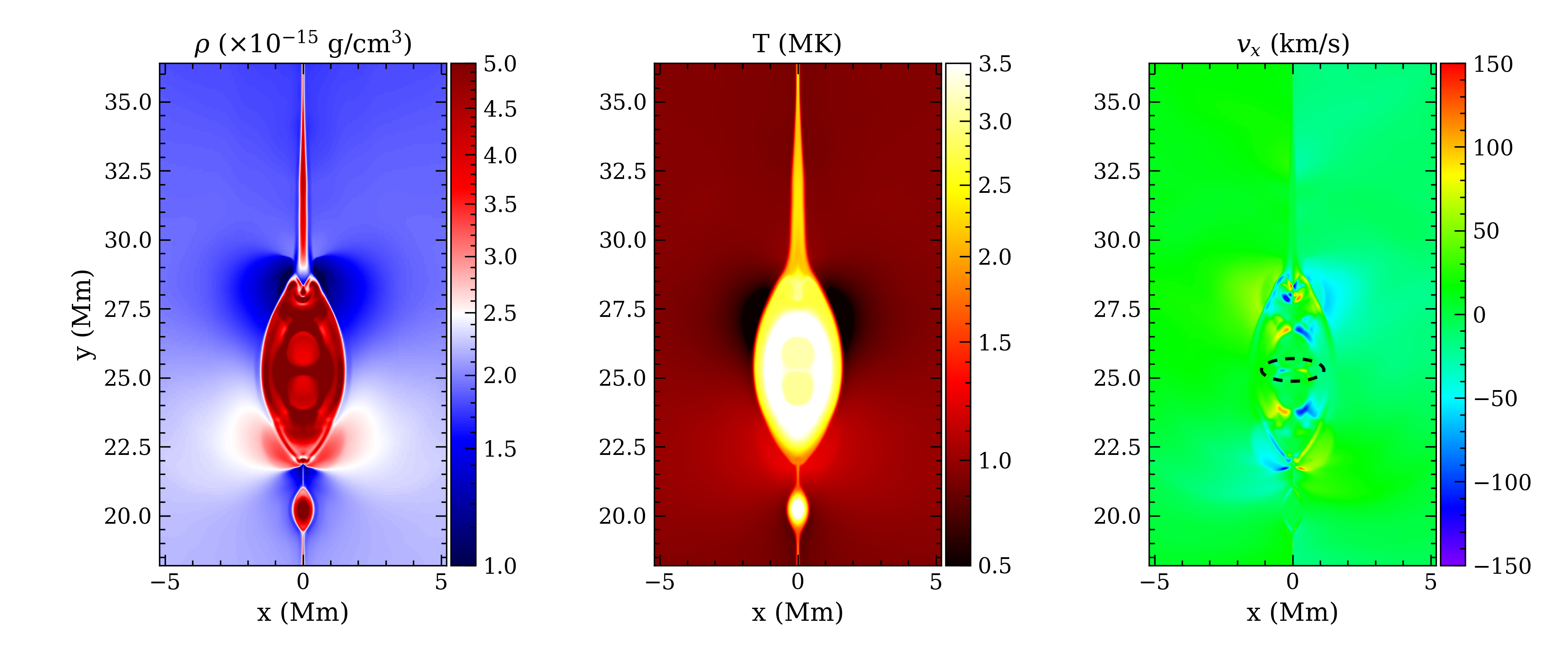}
    \caption{Spatial variations of the plasma density ($\rho$), temperature ($T$), and horizontal velocity ($v_x$) are shown from left to right in each column, as obtained from the MHD simulation by SMI25. The temporal evolution of these quantities at $t = 118.02$, $128.81$, and $145.98$ s is presented in the top, middle, and bottom panels, respectively. The nanojet-like bi-directional outflows are apparent in the $v_x$ maps within the dashed ellipses in the top and middle panels, whereas these flows almost disappear within the marked ellipse in the bottom panel. The horizontal dashed line in the top-left panel denotes the $y$-level at 27.2 Mm. An animation of the time evolution is available online.}
    \label{fig:MHD}
\end{figure*}

The system is in mechanical equilibrium at the initial state. To trigger reconnection, localized, high-velocity pinches at $y=23.33$, 30, and 36.67 Mm, such that the velocity is higher the higher the location in the atmosphere. These perturbations initiate forced reconnections at those heights, and lead to the tearing instability of the CS, and to a gradual development of two primary flux ropes at $y\approx 27$ and 33 Mm. The spatial inhomogeneity of the vertical component of the Lorentz force drives a relative motion between the flux ropes. Figure~\ref{fig:MHD} shows the density, temperature, and horizontal velocity ($v_x$) at $t=118.02$, 128.81, and 145.98 s. Interaction between the flux ropes begins at $t\gtrsim100$~s (see the animation associated with Figure~\ref{fig:MHD}). Opposite-polarity field lines reconnect at the interface of those flux ropes, generating bidirectional flows as marked by the ellipse at the top-right panel for $t=118$~s. The appearance of this flow remains present at an advanced stage at around $t=129$~s, when the upper flux rope starts to merge with the lower one, which leads to a transition from a purely 2D to 2.5D (or component) reconnection. By $t=146$ s, $v_x$ flows nearly disappear within the ellipse (see the bottom-right panel) as the guide field dominates. These bi-directional ejections have a total energy of $\sim3.3\times10^{24}$ erg, which is consistent with nanoflare-scale energy, and share similarities with the dynamic, and thermodynamic properties of the observed nanojets as described in SMI25. 

\begin{figure*}[hbt!]
    \centering
    \includegraphics[width=0.25\linewidth]{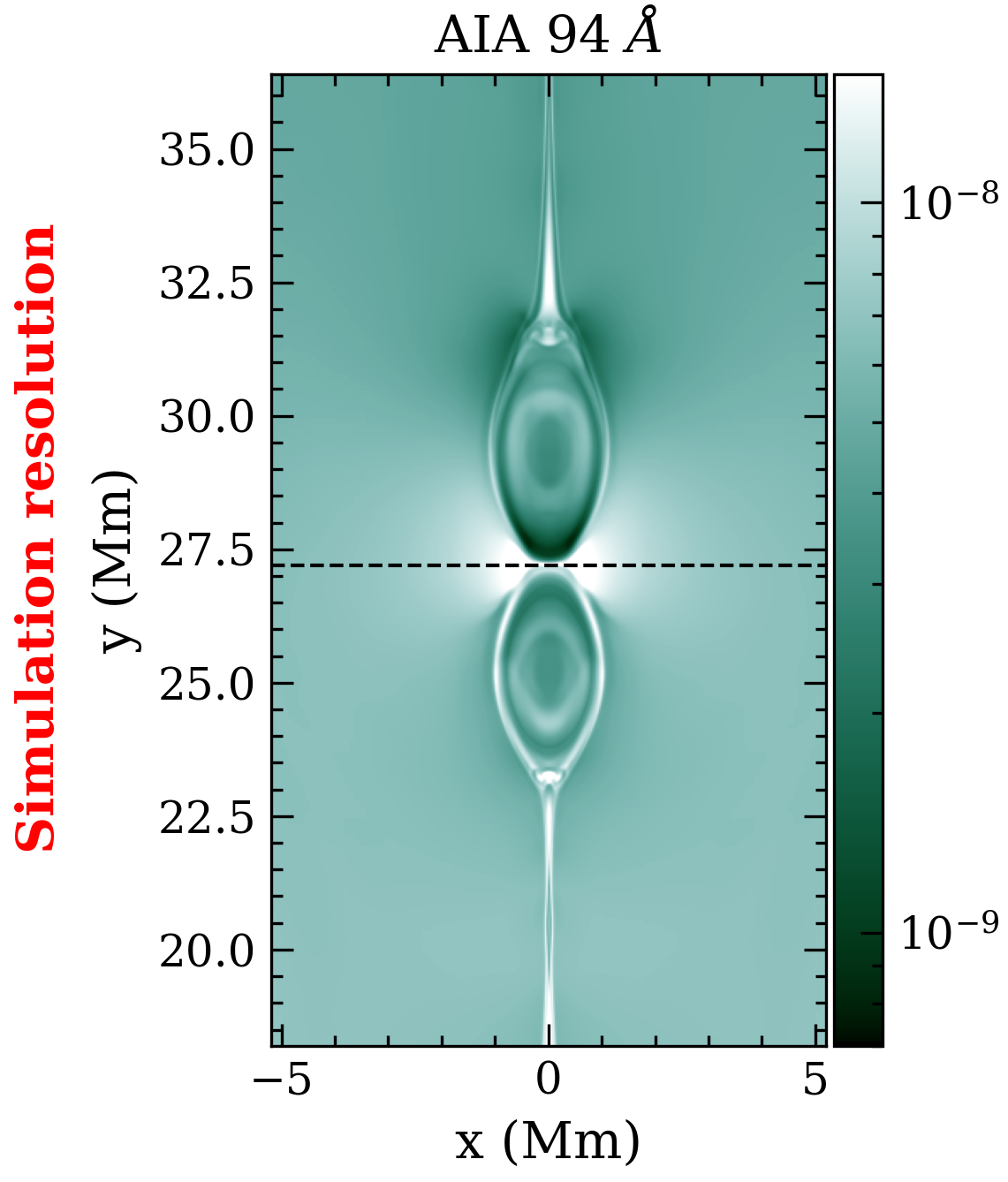}
    \includegraphics[width=0.23\linewidth]{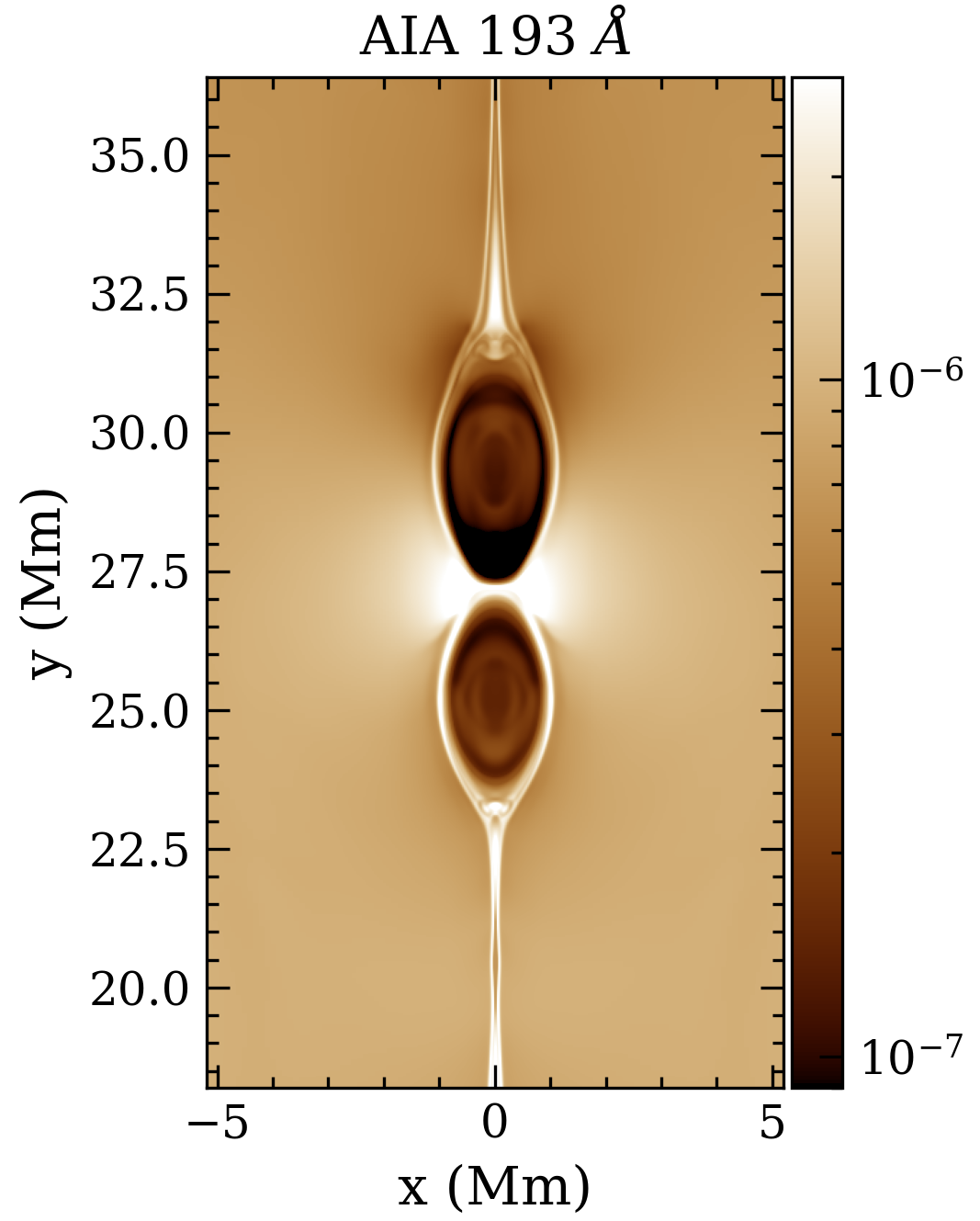}
    \includegraphics[width=0.23\linewidth]{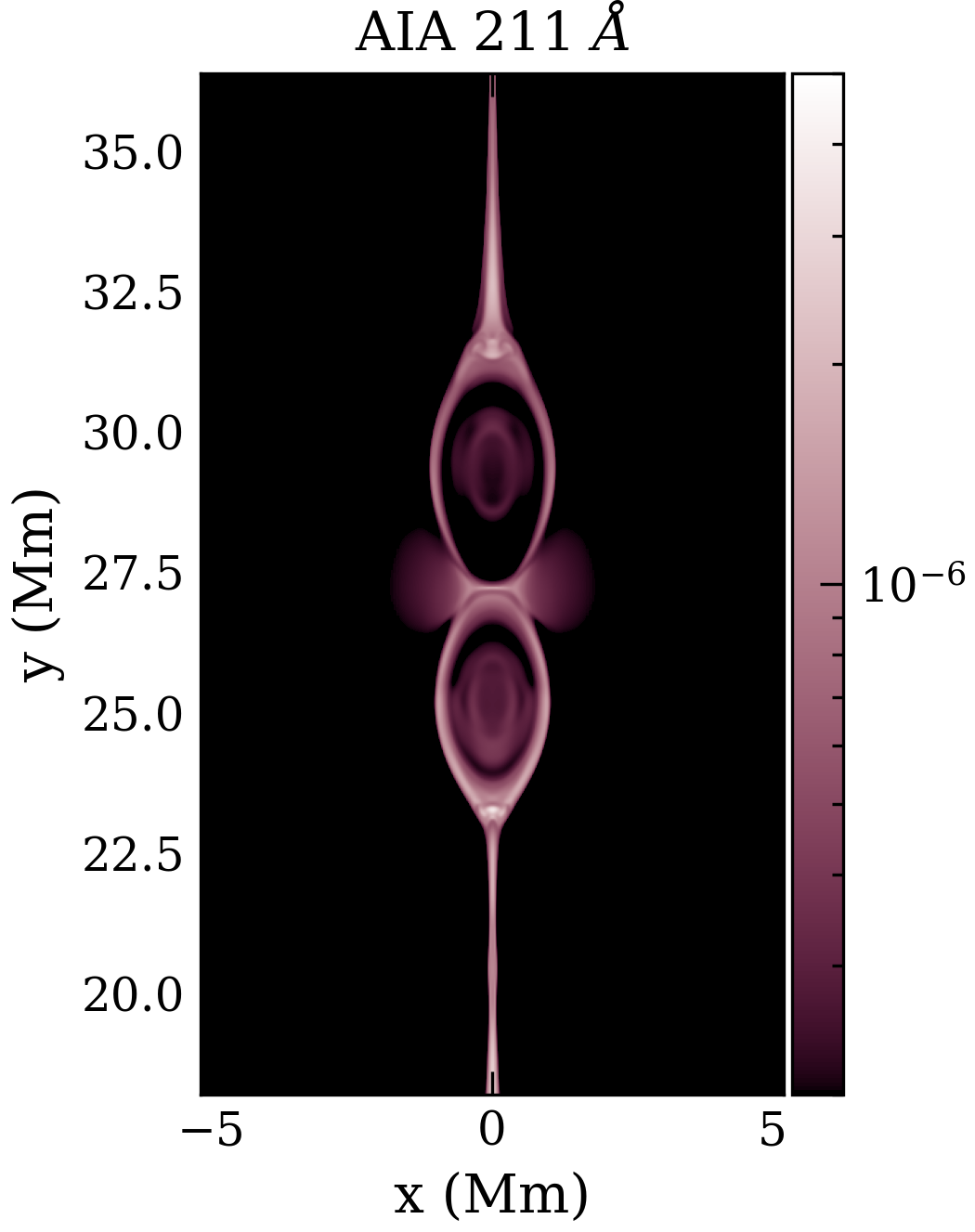}
    \includegraphics[width=0.25\linewidth]{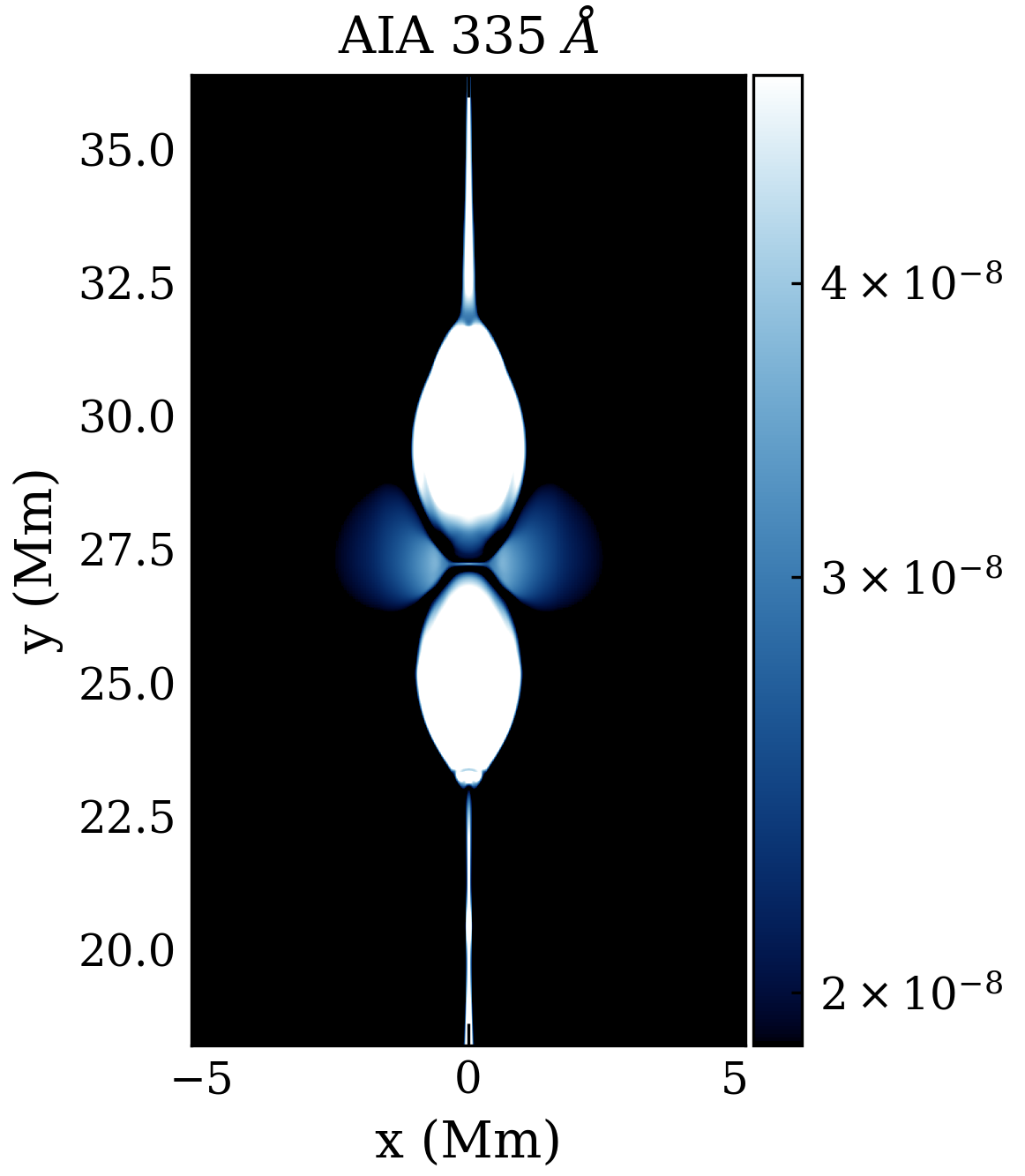}

    \includegraphics[width=0.25\linewidth]{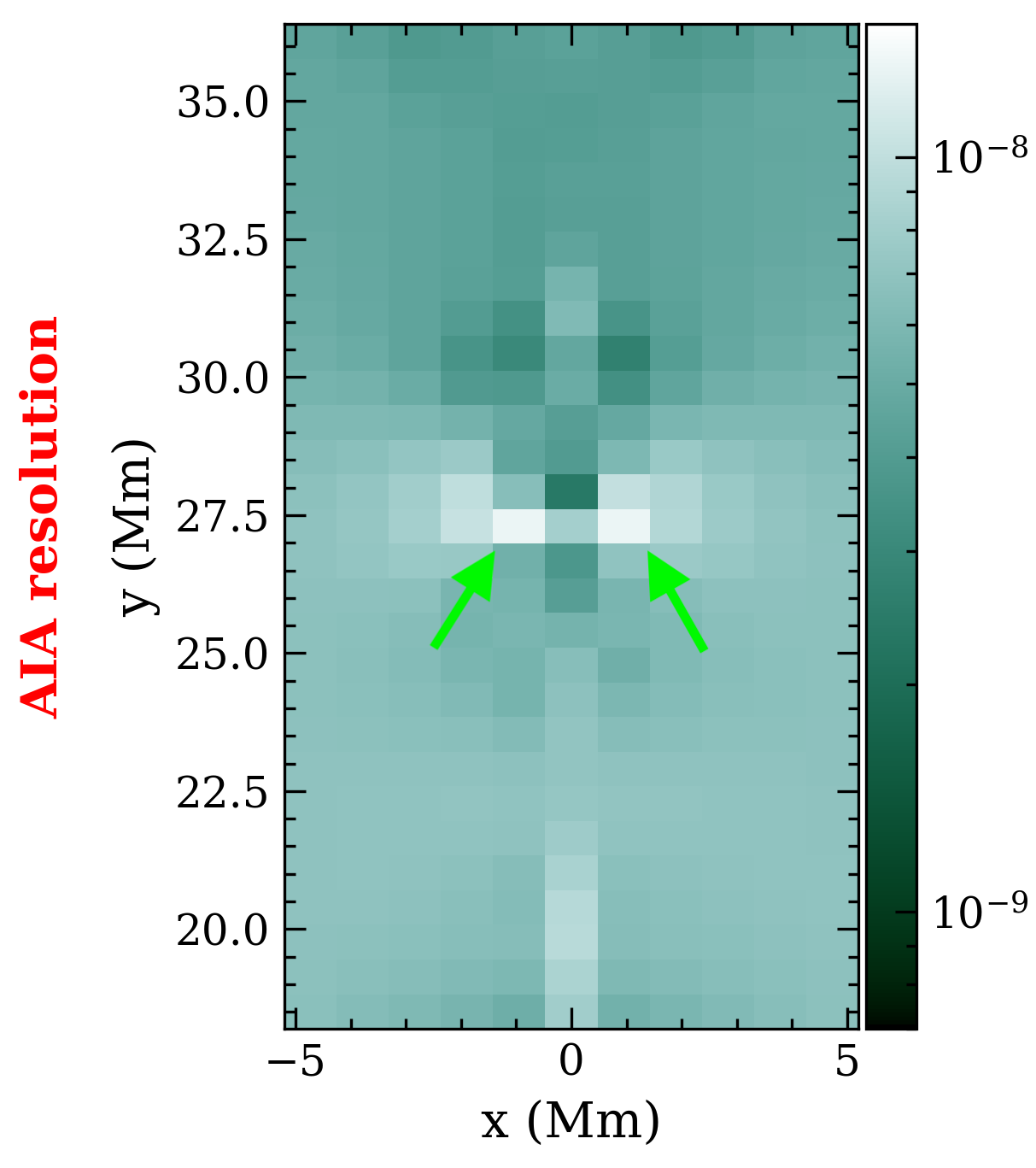}
    \includegraphics[width=0.23\linewidth]{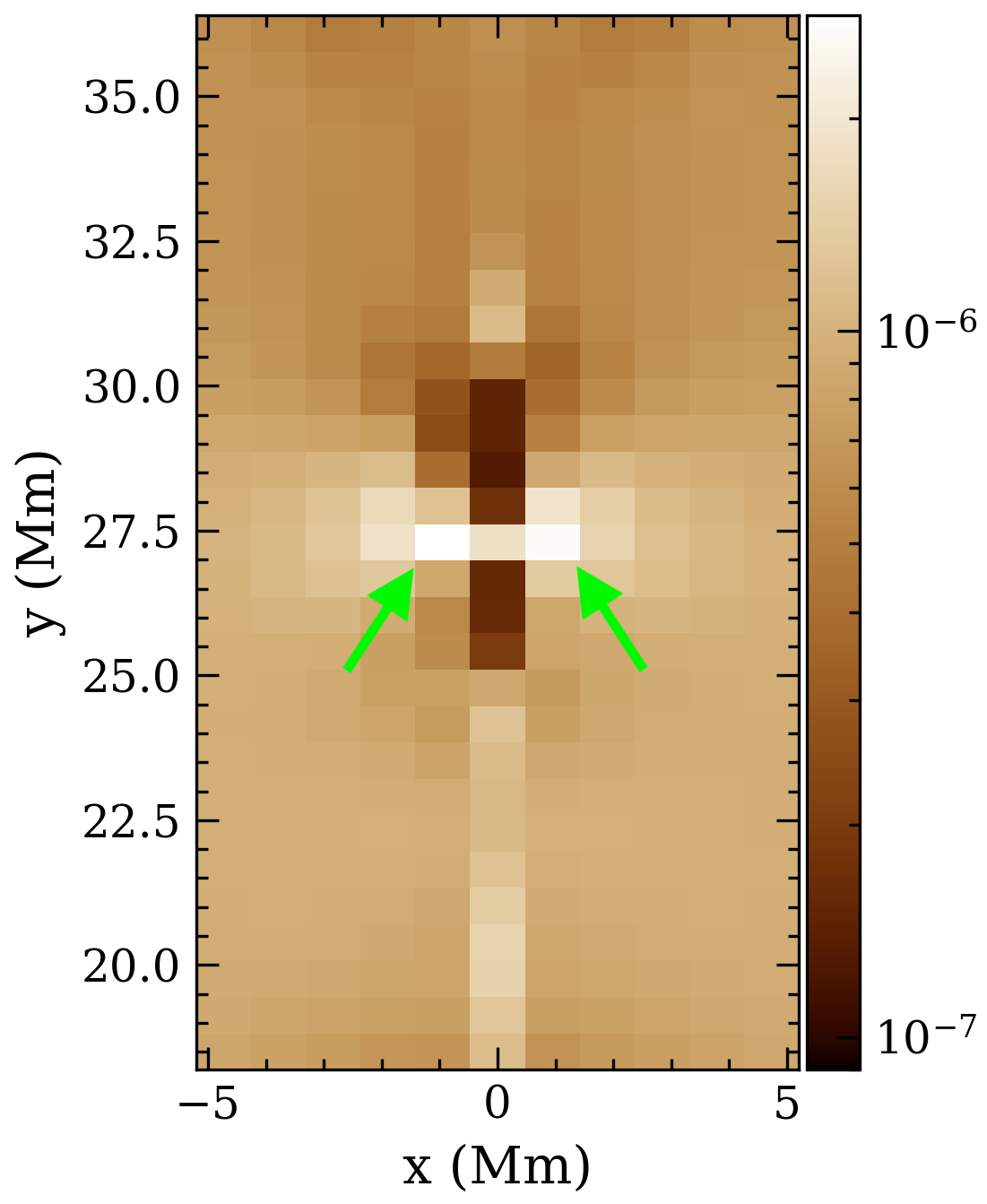}
    \includegraphics[width=0.23\linewidth]{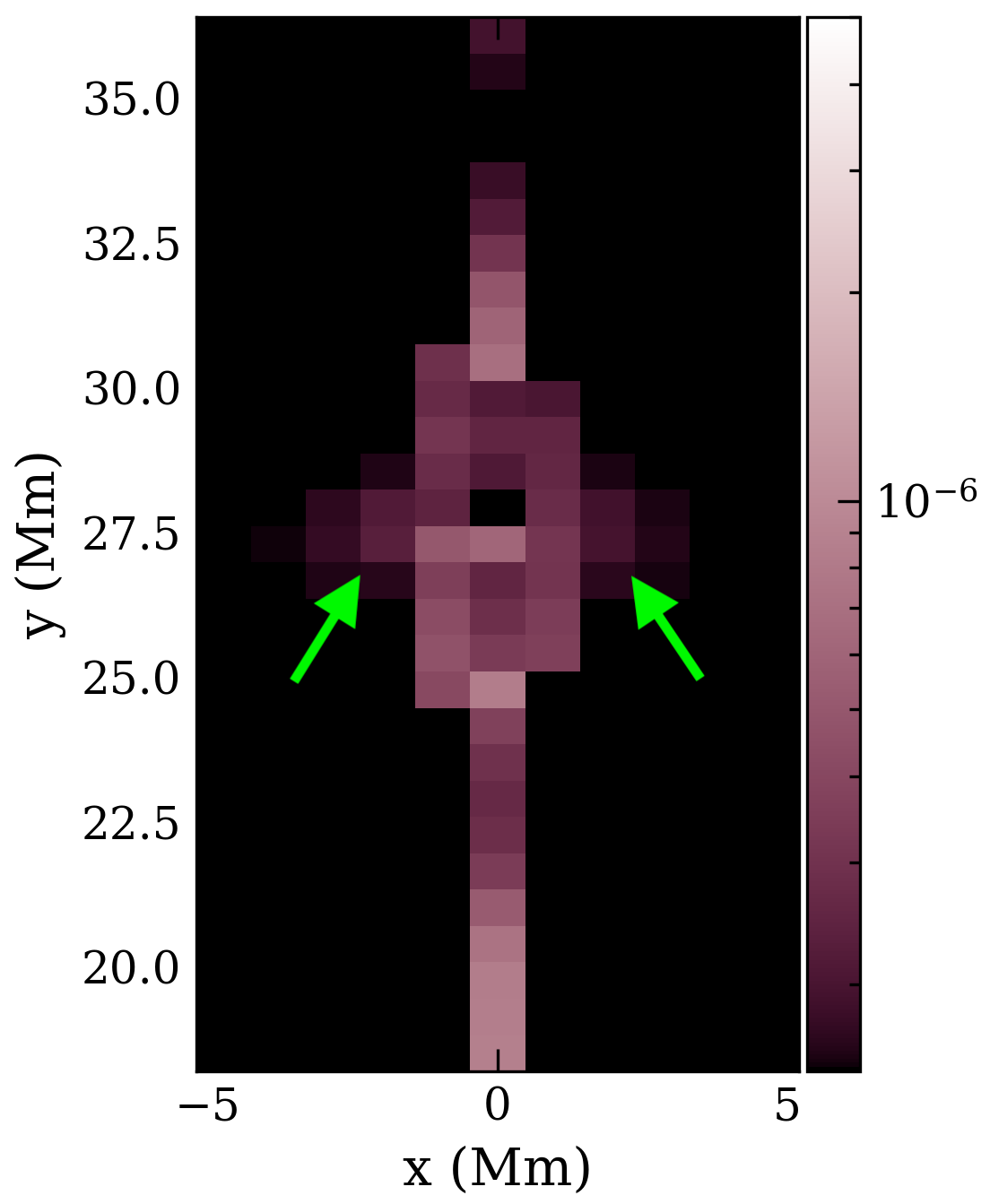}
    \includegraphics[width=0.25\linewidth]{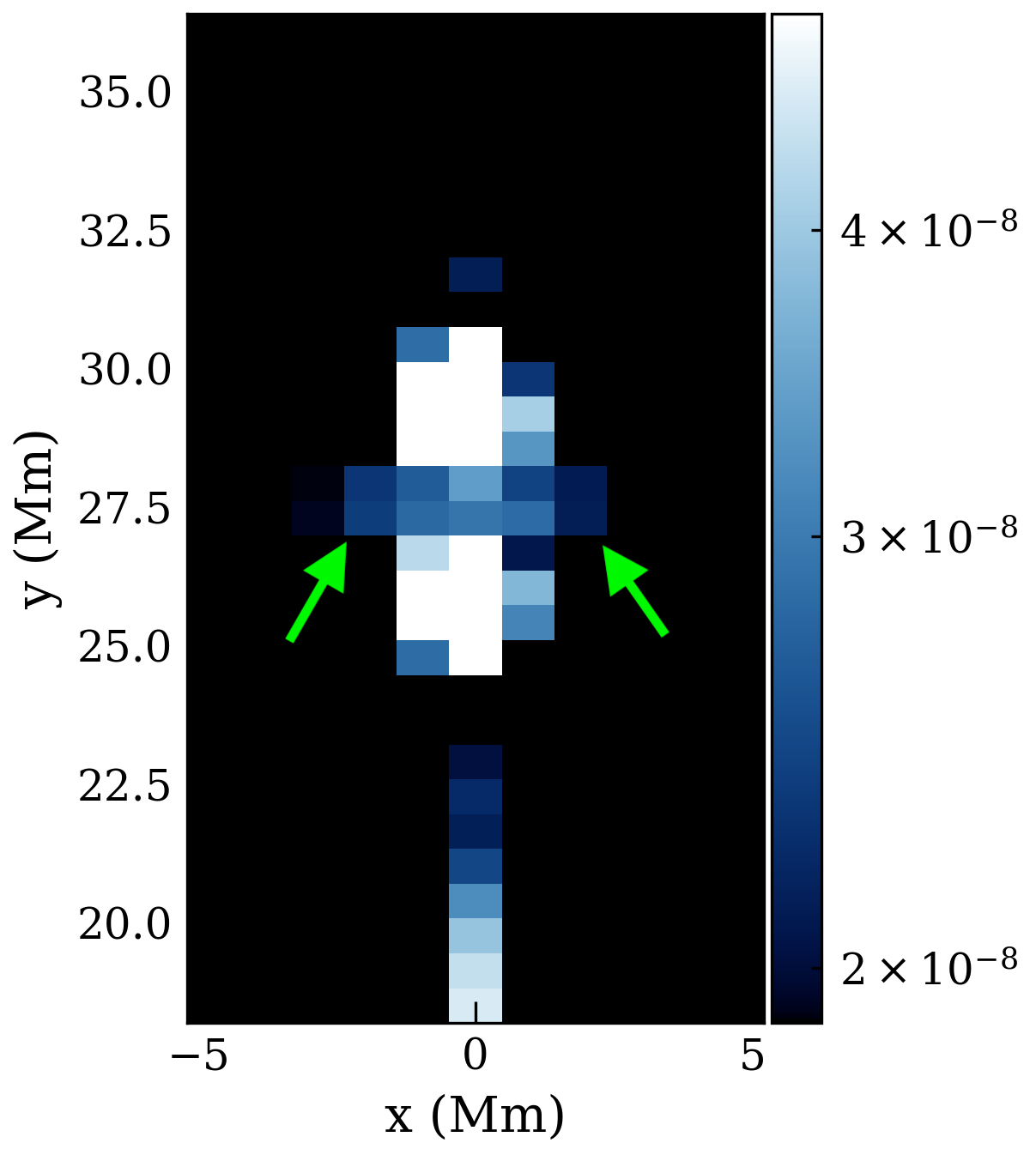}
    \caption{Top row: Synthetic emissivity maps in various SDO/AIA broadband channels at the spatial resolution of the simulation at $t = 118.02$~s. The emissivity values, given in DN cm$^{-1}$ s$^{-1}$ pix$^{-1}$, are indicated by the corresponding color bars. The black dashed line in the left panel marks the same location as in the top-left panel of Fig.~\ref{fig:MHD}. Bottom row: Same as the top row, but degraded to a spatial resolution of $1''.2$ comparable to AIA resolution, where the marked green arrows show the location of the nanojet-like feature. An animation of the time evolution is available online.}
    \label{fig:aia_synthesis}
\end{figure*}

\begin{figure*}[hbt] 
    \centering
    \includegraphics[width=0.8\linewidth]{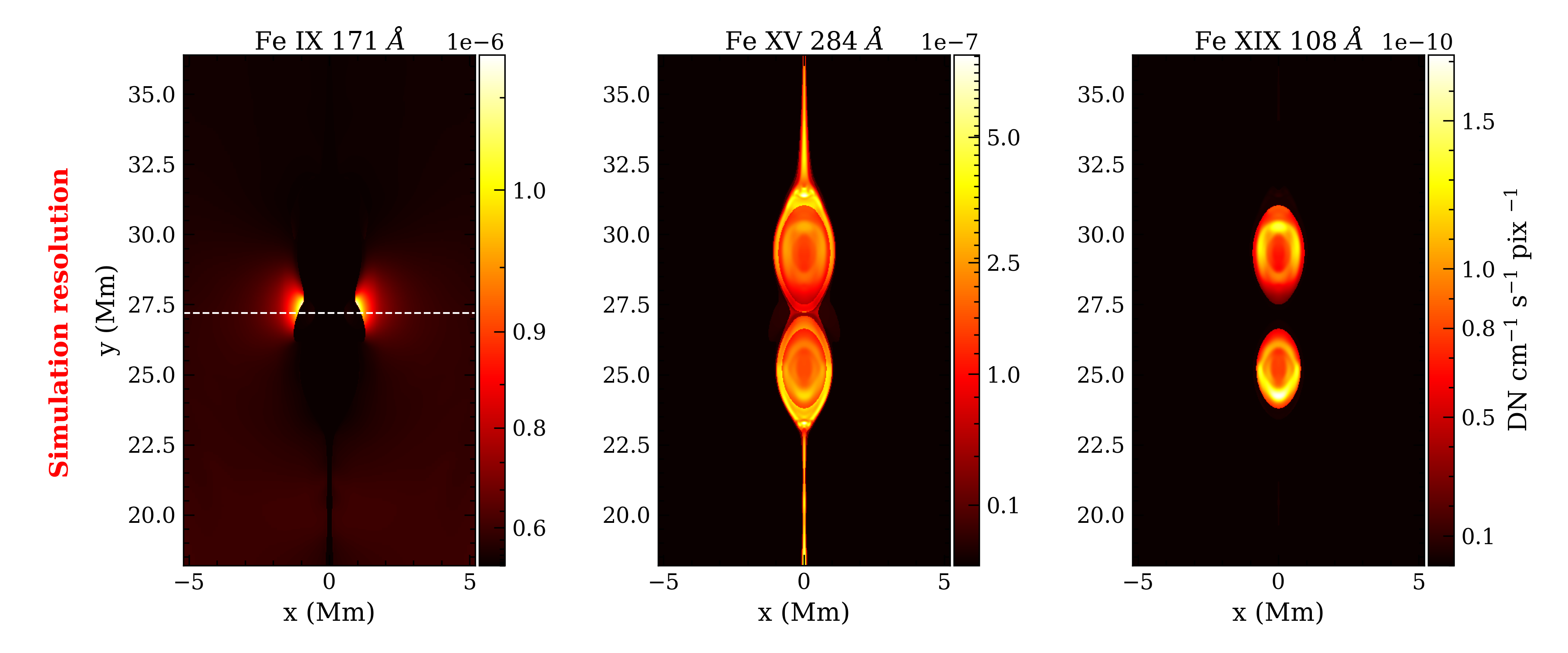}
    \includegraphics[width=0.8\linewidth]{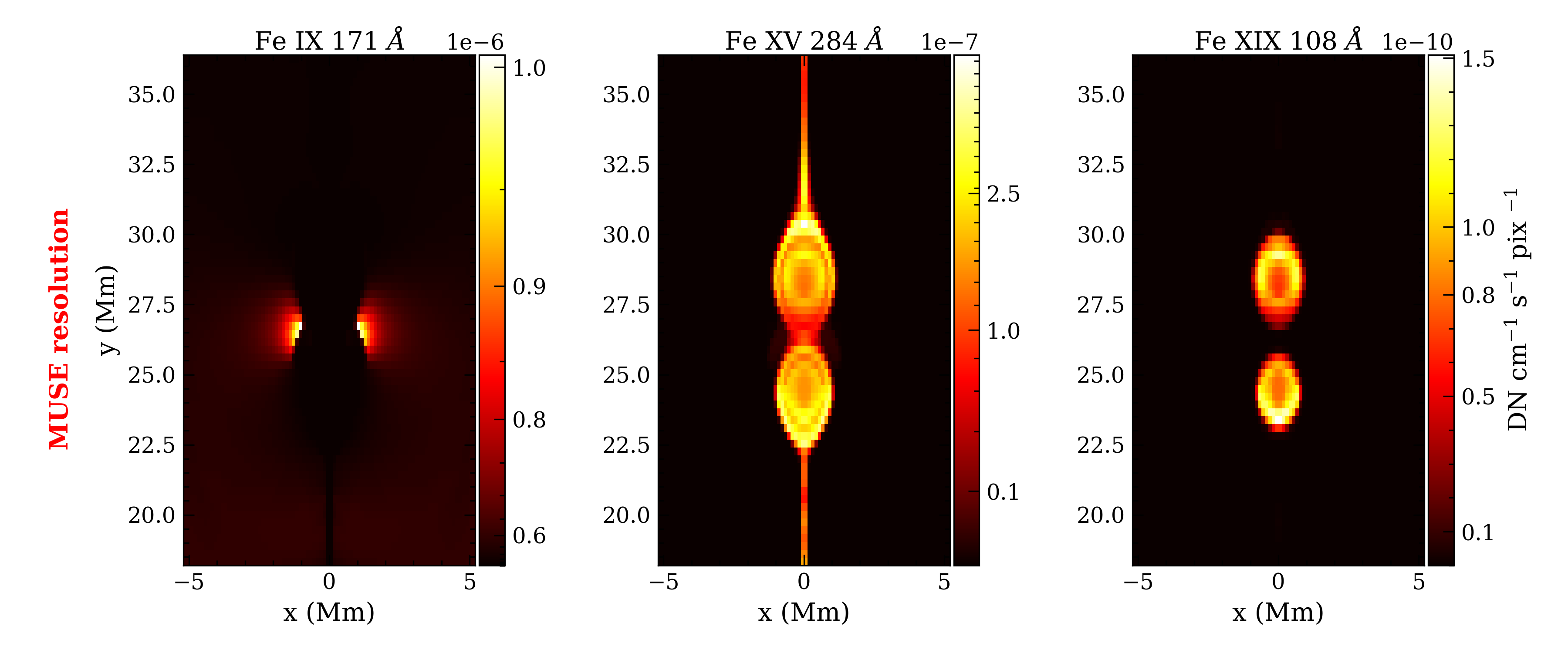}
    \caption{Top row: synthetic emissivity maps in different MUSE lines with the spatial resolution of the simulation at $t=118.02$~s, where the horizontal dashed line is a marker at $y=27.2$~Mm. Bottom row: same as the top row, but with a resolution of $0''.4$, and $0''.167$ along the $x$ and $y$ directions respectively, compatible to MUSE.}
    \label{fig:muse_synthesis}
\end{figure*}

\begin{figure*}[hbt] 
    \centering
    \includegraphics[width=0.3\linewidth]{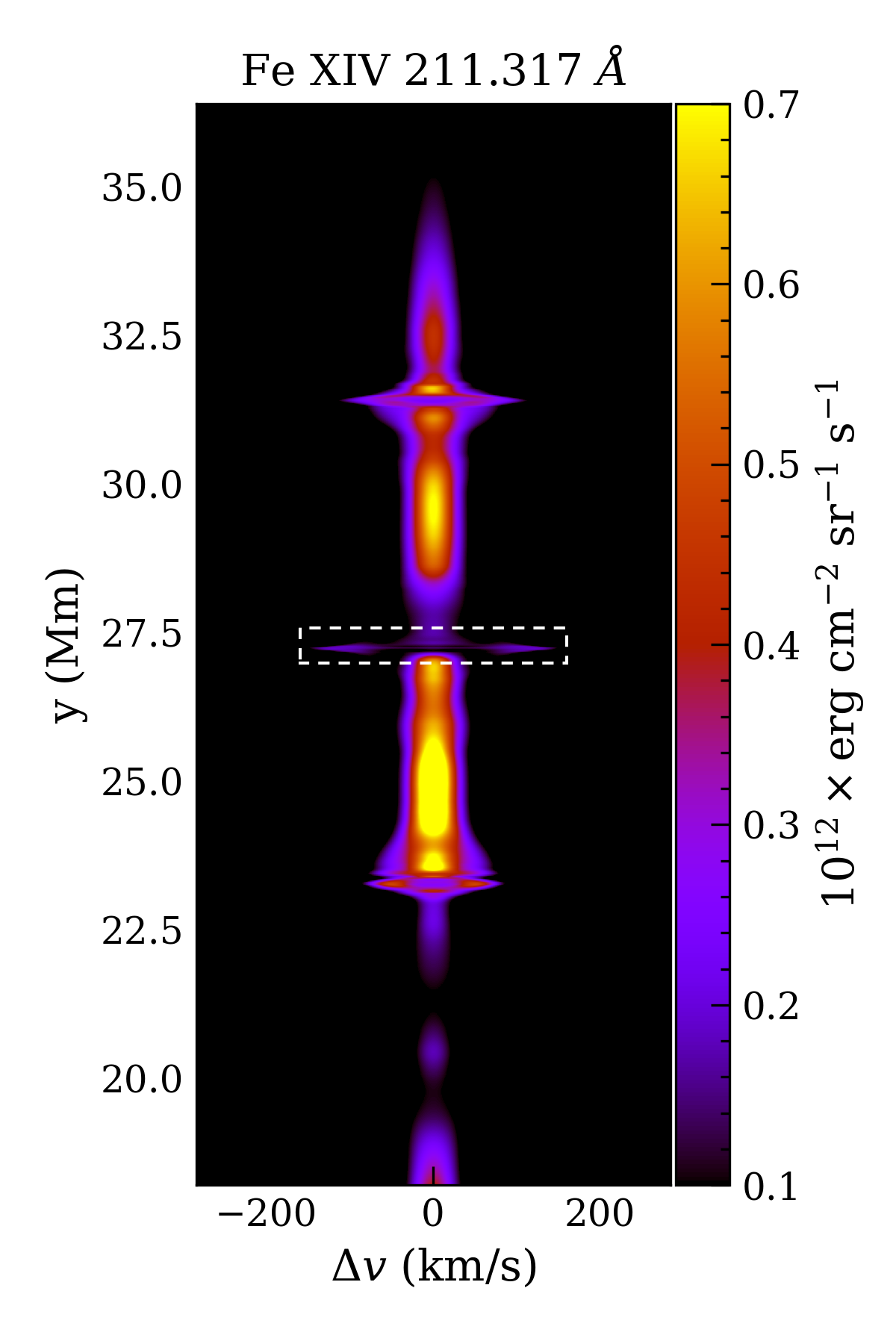}
    \includegraphics[width=0.3\linewidth]{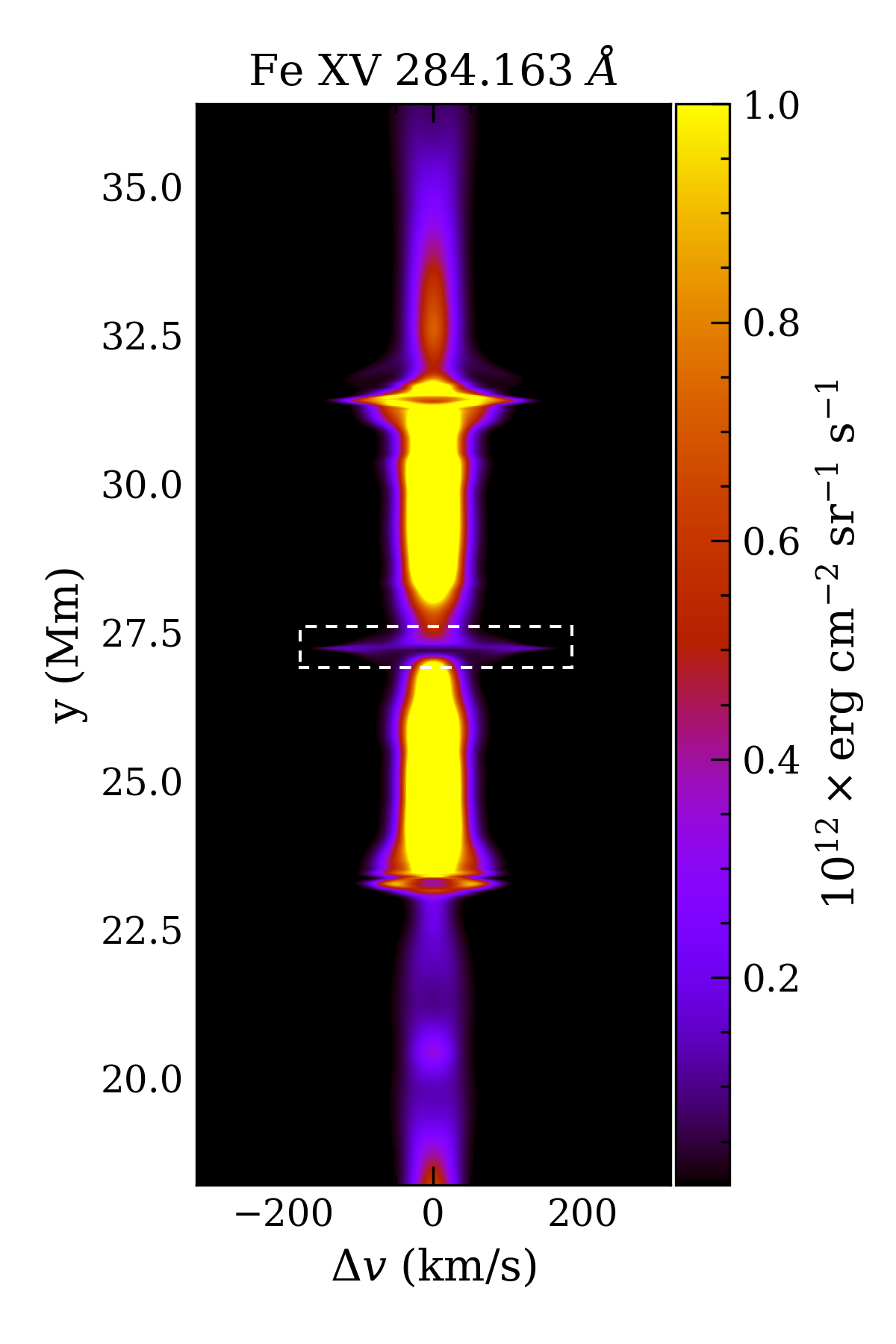}
    \includegraphics[width=0.3\linewidth]{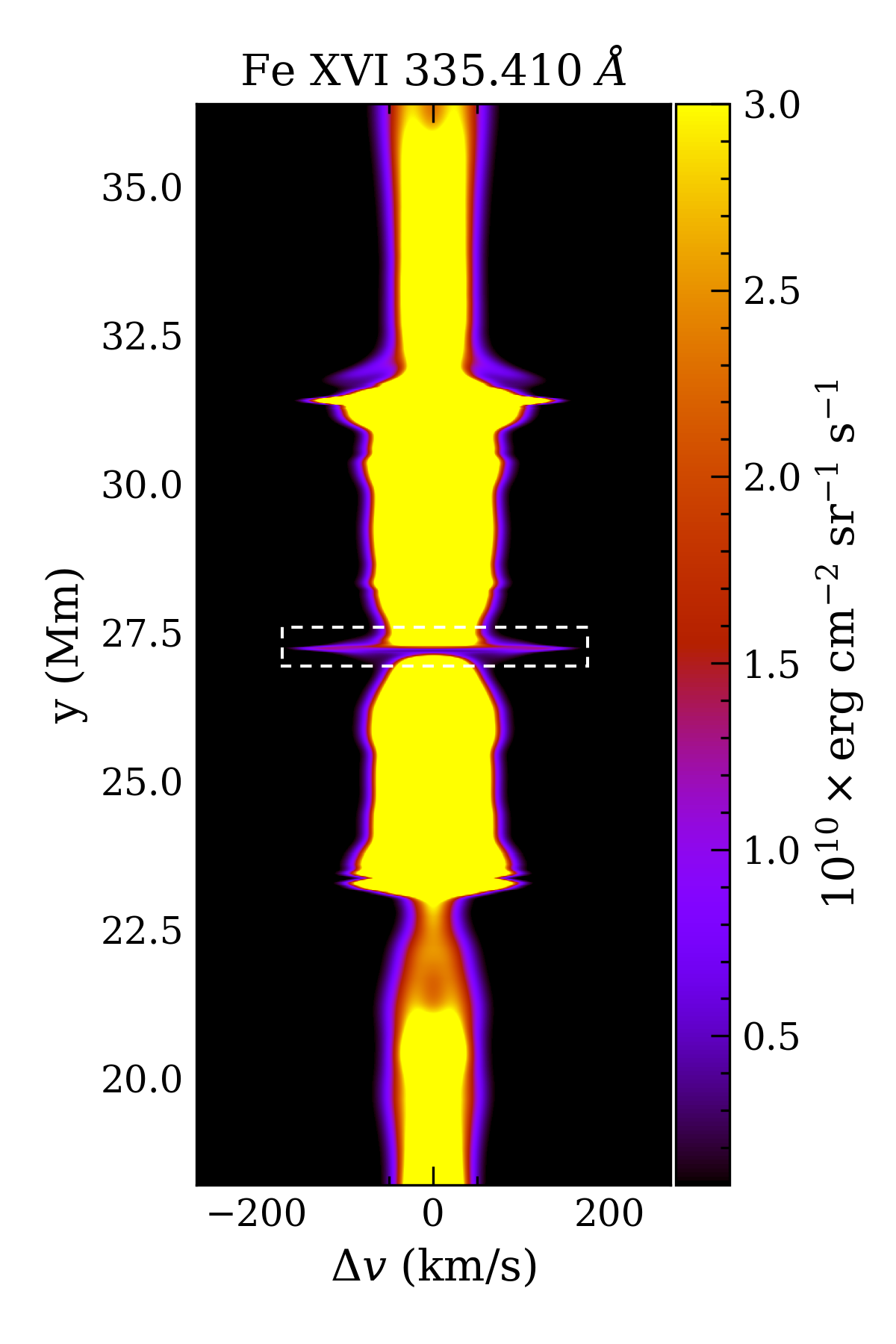}
    \includegraphics[width=0.5\linewidth]{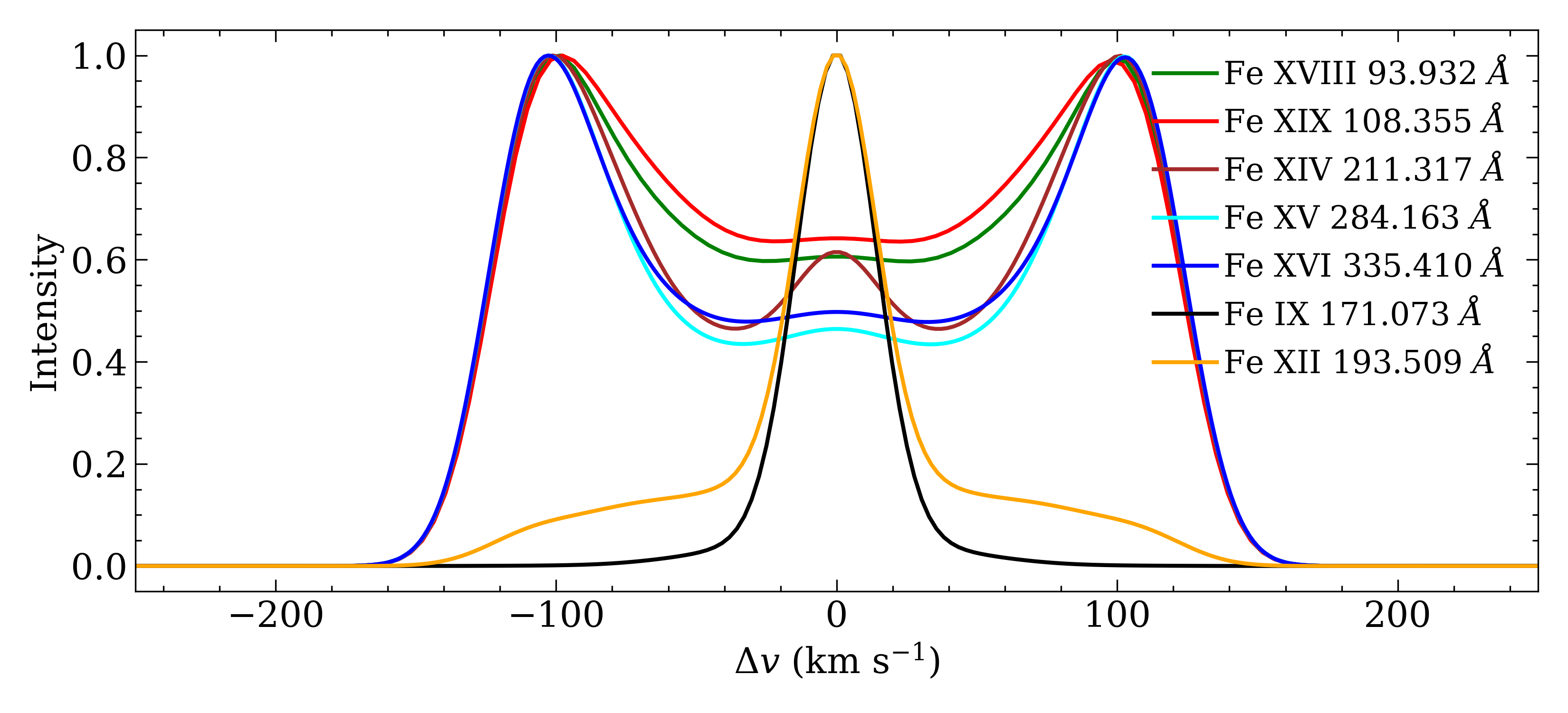}
    \includegraphics[width=0.35\linewidth]{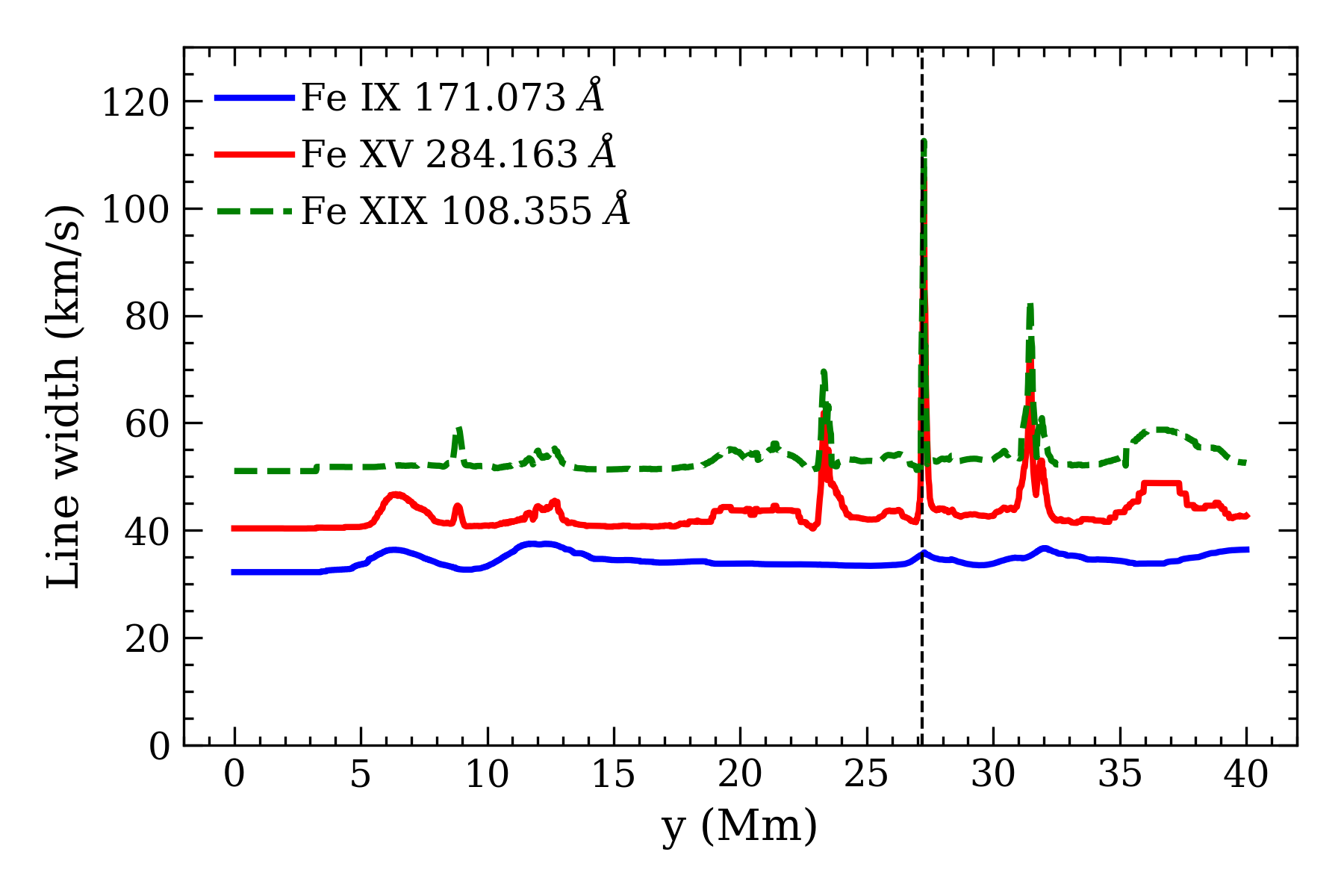}
    \includegraphics[width=0.32\linewidth]{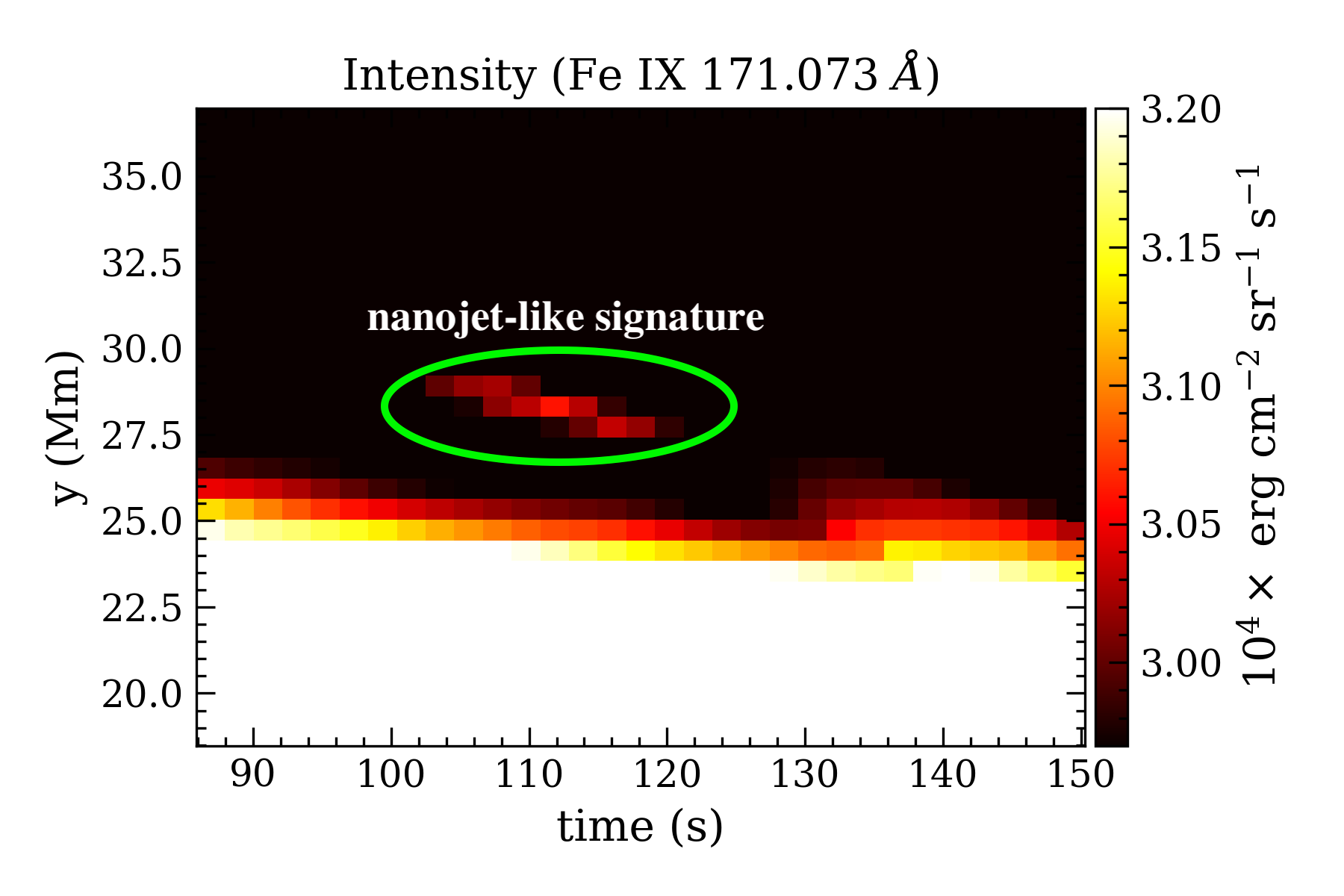}
    \includegraphics[width=0.32\linewidth]{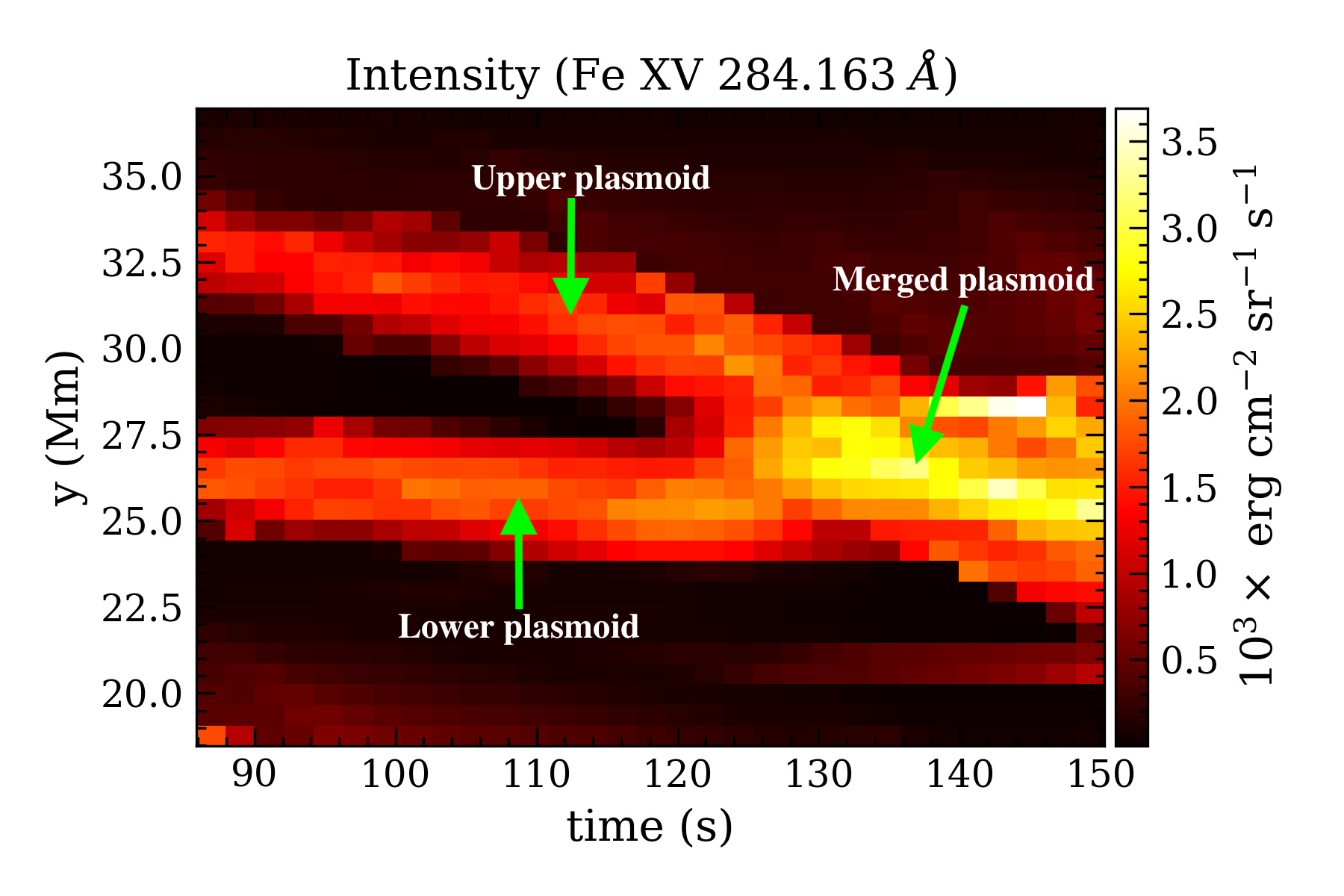}
    \includegraphics[width=0.32\linewidth]{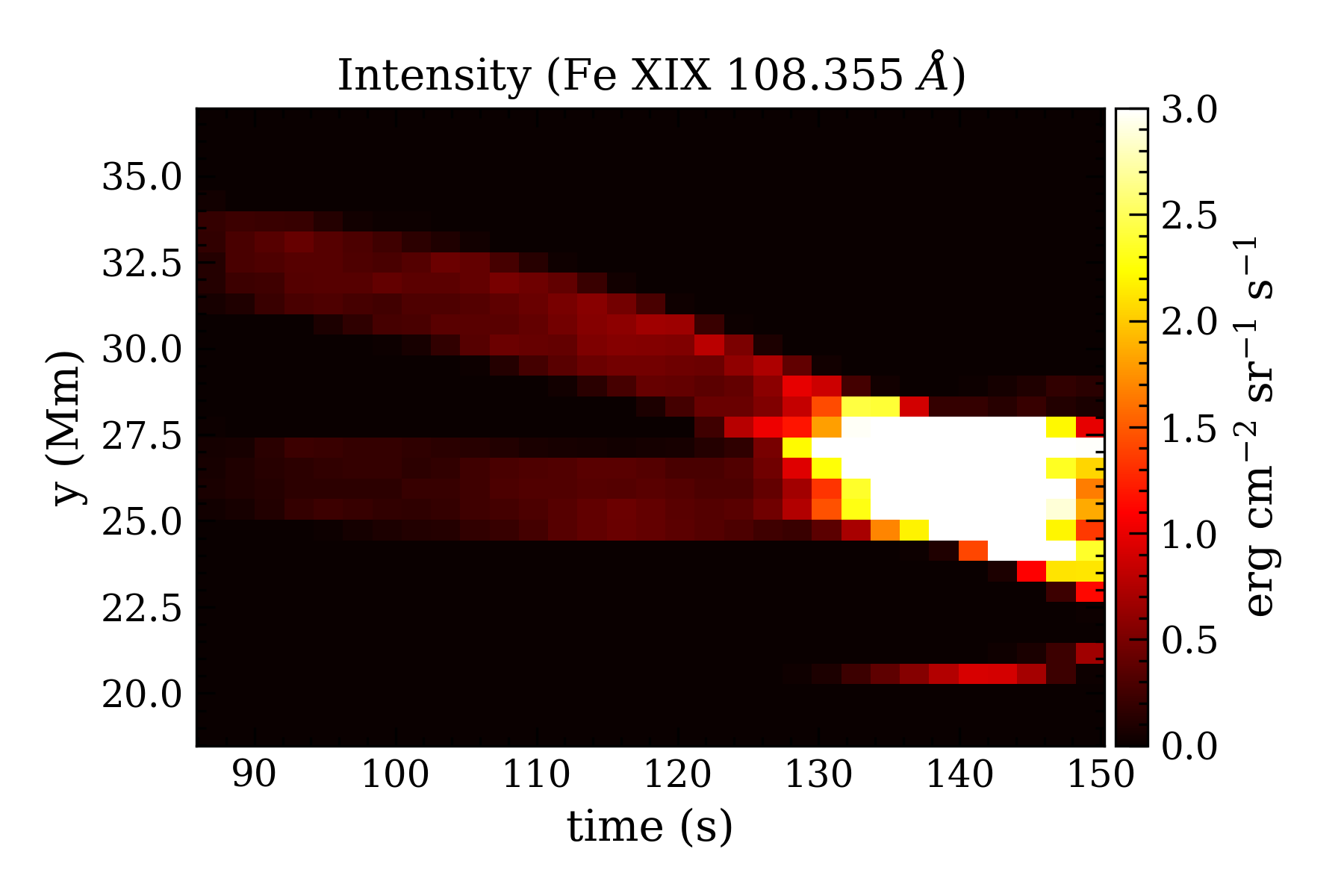}
    \caption{Top row: Spectral maps in different emission lines at $t=118.02$~s, where the intensities are calculated by integrating along the $x$ direction. The white boxes are placed at around $y=27.2$~Mm to highlight spectral signature of the nanojet-like ejections. Middle row: The left panel represents spectral profiles for the emission lines (marked with the corresponding legends) at $y=27.2$~Mm, and the right panel shows the variation of total line width along the $y$ direction, where the vertical dashed line is a marker at $y=27.2$~Mm. Bottom row: Time-distance map of the intensities, which are obtained by integrating along the slits, which are placed horizontally (along $x$) at 31 different $y$-levels. The green ellipse in the left panel shows the region with brightening between $y\approx 27$ and 30 Mm and $t\approx 100$ and 120 s, showcasing the nanojet-like signature. The maps at the middle and right panels highlight the evolution of the merging plasmoids (as marked in the middle panel).}
    \label{fig:synthetic_spectra}
\end{figure*} 

\section{Forward modeling results and discussion}\label{sec:results}

\subsection{Synthetic observation}\label{subsec:synthesis}
To obtain synthetic observations in different EUV channels, we perform forward modeling in the optically-thin approximation, relevant for a typical coronal medium. The detailed methodology is described in Appendix~\ref{sec:forward modeling}.

To showcase the imaging signatures of nanojet-like ejections in various EUV channels, we generate synthetic emissivity maps in the 94, 193, 211, and 335~\AA\ passbands of AIA. In addition, we construct synthetic emissivity maps for the Fe IX 171, Fe XV 284, and Fe XIX 108~\AA\ lines, which are relevant to the upcoming MUSE mission. The effective pixel areas for AIA channels and MUSE lines are incorporated in the respective contribution functions for estimating the emissivity in DN cm$^{-1}$ s$^{-1}$ pix$^{-1}$ units. We also synthesize spectral line profiles for Fe XII (193~\AA), Fe XVIII (94~\AA), Fe XIV (211~\AA), Fe XVI (335~\AA), Fe IX (171~\AA), Fe XV (284~\AA), and Fe XIX (108~\AA) ions.   

In the top and bottom rows of Fig.~\ref{fig:aia_synthesis}, we show the emissivity maps at $t=118.02$~s in various AIA passbands to highlight the imaging signature of the nanojet-like ejection. The maps in the top row are rendered at the highest simulation resolution of 26 km in both directions. An enhancement of emissivity is evident in a localized region near $y=27.2$~Mm (marked by the black dashed line in the top-left panel), extended horizontally to the left and right from $x=0$. This feature is co-spatial with the region of enhanced density apparent in the top-left panel of Fig.~\ref{fig:MHD}, which coincides with the jet-like high-velocity features (up to 150 km s$^{-1}$) in the top-right panel in the same figure. The bottom row of Figure~\ref{fig:aia_synthesis} presents the emissivity maps degraded to $1.2''$ ($\approx 870$~km) resolution, comparable to AIA. The nanojet-like signatures are still discernible at the same height, as indicated by the green arrows, but they appear much fainter compared to the counterpart maps with simulation resolution. These signatures of the nanojets extend up to 3 pixels away from the CS location, corresponding to a length of $\lesssim 2.6$~Mm. We estimate the temperature of these features to be $\lesssim 1.5$~MK from the simulation, whereas the upper and lower plasmoids reach $\approx 3.2$~MK (see top-middle panel of Figure~\ref{fig:MHD}). The dominant line contributing to the AIA 335~\AA\ passband has a formation temperature of $\approx 2.8$~MK, leading to brightening at the plasmoid regions in the 335~\AA\ emissivity maps. In contrast, the brightenings are much fainter in the other AIA passbands. 

Figure~\ref{fig:muse_synthesis} presents the corresponding synthetic emissivity maps for the MUSE lines, also at $t=118.02$~s. The top row shows the maps at the original simulation resolution, while the bottom row displays the maps degraded to the MUSE slit width and pixel size of $0''.4$ along the $x$ direction and $0''.167$ along the $y$ direction, respectively. The Fe IX 171~\AA\ line, which peaks near $0.9$~MK, is more closely aligned with the temperature of the nanojet-like features around $y=27.2$~Mm than the higher-temperature lines forming at $2.5$~MK (Fe XV 284~\AA) and $10$~MK (Fe XIX~108\AA). As a result, the Fe IX (171~\AA) line corresponds to the strongest signal for the nanojet-like feature. In contrast, the current sheet and plasmoids exhibit temperatures of $\approx 2.5$~MK and $\approx 3.2$~MK, respectively. Accordingly, the Fe XV 284~\AA\ emissivity map highlights both the current sheet and plasmoid structures, whereas the Fe XIX 108~\AA\ map highlights the signal only for the plasmoid region. To measure the prediction of the photon counts associated with the nanojet-like feature shown in the emissivity maps in Figs.~\ref{fig:aia_synthesis} and \ref{fig:muse_synthesis}, we assume that the feature extends by $1$~Mm along the invariant direction, which is comparable to its width along the $y$-direction. This leads to the maximum intensity at a height of $y=27.2$~Mm is $\approx 550$ in the AIA~193~\AA\ and AIA~211~\AA\ channels, and $\approx100$~DN s$^{-1}$ pix$^{-1}$ in the MUSE~Fe~IX~(171~\AA) line. In contrast, the intensity in the other channels and spectra lines has significantly lower values, namely $\lesssim 10$~DN s$^{-1}$ pix$^{-1}$. However, it is worthy to mention that we do not introduce any readout or Poisson noises when producing the synthetic images for the AIA channels and the MUSE lines. Similar analysis without any noise has been done for small-scale energetic events, e.g., nanoflare storms \citep{Cozzo:2024, Johnston:2025}, and also for the microphysical processes, e.g., MHD turbulence \citep{Sen:2021, Shen:2022NatAs}. Nevertheless, the extent to which the inclusion of additional instrumental effects (noise), the presence of increased material along the line-of-sight \citep{Ernest:2025}, or an enhanced integration time may smooth or smear out the nanojet signatures remains an important topic for future investigation.

In Fig.~\ref{fig:synthetic_spectra}, the top row shows spectral maps for different iron emission lines as a function of height, $y$, with the spectral dimension given in abscissas. The intensities at each height are calculated through line of sight (LOS) integration along $x$, and with a spectral resolution of $15$~m\AA. The white boxes in each panel, located near $y=27.2$~Mm, highlight remarkable spiky features along the $\Delta v$ axis. These features indicate that the flows are collimated along the $x$ direction. The peak intensities of these spikes (near $y=27.2$~Mm) are shifted to the left and right of the $\Delta v$ axis center, demonstrating the bi-directional nature of the flows. The left–right symmetry of these spikes reflects the corresponding symmetry in the velocity, density, and temperature distribution of the ejections, also evident in the top row of Figure \ref{fig:MHD}. There are other notable spikes at $y\approx 32$ and 23~Mm present in all the spectral maps. The former arises due to the plasma inflows caused by localized plasma pressure gradients, and the latter arises due to tiny reconnection outflows, as seen in the top-right panel of Figure \ref{fig:MHD}. These synthetic spectral maps represent example outputs similar to the spectrographs that we can obtain from the current and upcoming telescopes. For instance, a Fe XIV 211.32~\AA\ line is available in the Hinode / \textit{EUV Imaging Spectrometer} \citep[EIS;][]{Culhane:2007} spectrograph which consists of two modes with different widths of $1''$ and $2''$ with spectral resolution of $22$~m\AA\ for each slit \citep{Culhane:2007}, and in the upcoming Solar-C / Extreme Ultraviolet High-Throughput Spectroscopic Telescope \citep[EUVST;][]{EUVST:2021} mission which is expected to contain a single slit with a higher spatial resolution of $0''.4$ and spectral resolution between $17–21.5$~m\AA. The spectrograph for the Fe XV (284.163~\AA) emission line will be present in the upcoming MUSE mission, which will contain 35 slits with spatial resolution $\lesssim~0''.4$ \citep{MUSE-Depontieu:2020}. The prediction of such synthetic spectral maps for the MUSE lines is also reported in \cite{DePontieu:2022}, and \cite{Cozzo:2024}. The spectral map shown for the Fe XVI (335.410~\AA) emission line also showcases the signature of Doppler shift at $y=27.2$~Mm, and underscores the importance for future instrumentation as there are no available spectrographs that can observe this line in any current (or near-future) observatories.       

To appreciate the spectral profiles for the nanojet ejection, we extract a cut at $y=27.2$~Mm from the spectral maps (with intensities normalized to unity), as shown in the left panel of the middle row of Figure \ref{fig:synthetic_spectra}. For Fe XIV, XV, XVI, XVIII, and XIX, we find two distinct peaks at $\Delta v \approx \pm 100$ km s$^{-1}$, corresponding to Doppler shifts in the red and blue wings from their central wavelengths at 211.317, 284.163, 335.410, 93.932, and 108.355 \AA\, respectively. These peaks result from a complex combination of density, temperature, line-of-sight (LOS) velocity, spectral response function, and the contribution function of each line. In contrast, the Fe IX (171 \AA) and Fe XII (193 \AA) profiles peak at $\Delta v=0$ and exhibit no significant Doppler shifts compared to the nanojets' velocity ($\approx \pm100$~km s$^{-1}$). The Fe XV, XVI, XVIII, and XIX profiles are nearly flat around $\Delta v=0$, while the Fe XIV profile (brown curve) exhibits a small local peak at $\Delta v=0$. This feature reflects the dominant contributions to the intensity from the pixels with very low velocities. The right panel of the middle row presents the second moment of the velocity ($v_x$) distribution (Equation \ref{eq:2nd_mom}) as a function of $y$ for the MUSE lines. This second moment serves as a proxy for the combined effects of thermal, non-thermal, and instrumental broadening. The thermal broadening (see Eq.~\ref{eq:vth}) for each MUSE line using their respective peak formation temperatures, are 16 km s$^{-1}$, $27$~km s$^{-1}$, and $54$~km s$^{-1}$ for Fe IX, XV, and XIX lines, respectively. The instrumental broadening for each MUSE line is estimated by $\frac{2.9c}{\sqrt{8~ln2}} \frac{\Delta \lambda}{\lambda_f}$ \citep{2024A&A...687A.171F}, where, $\Delta \lambda=15$~m\AA\ is the spectral sampling and $\lambda_f$ is the wavelength for the $f$-th instrumental channel. This yields the instrumental line widths of 32.3, 19.5, and 51.2 km s$^{-1}$ for the Fe IX (171~\AA), Fe XV (284~\AA), and Fe XIX (108~\AA) lines, respectively. For Fe XV and XIX, the total line width peaks at $y=27.2$~Mm (at the nanojet-like feature height), with values of $\lesssim 100$~km s$^{-1}$. Two additional remarkable spikes are seen at $y\approx 23$~Mm and $y\approx 32$~Mm, corresponding to a velocity $\lesssim 80$~km s$^{-1}$. On the other hand, no notable spikes are present for the Fe IX line. 

\subsection{Time evolution}\label{subsec:time-evolution}

In the bottom row of Figure~\ref{fig:synthetic_spectra}, we present the space-time maps of intensity for the MUSE lines, where the intensities are calculated by integrating the pixels along $x$, and placing 31 slits at different $y$ levels between $y=18.2$ to 36.4 Mm with uniform spacing of $\approx 600$~km, in the time interval between $t=86$ and 150 s. 

The Fe IX (171~\AA) map highlights an intensity enhancement in a localized region (marked by a green ellipse) between $y \approx 27$–$29$~Mm and $t \approx 102$–$122$~s. This feature corresponds to collimated plasma ejection along the $x$-direction, intensity sensitive to the temperature of $\sim 1$~MK, and located in between the merging flux ropes. From the space–time map, we estimate the lifetime of this feature to be $\approx 20$~s. We identify this feature (within the green ellipse) at a space-time location where its brightening is not compromised by occultation from other portions of the merging plasmoids. At later stages of the plasmoid-merging process, this feature can no longer be clearly discerned in the space-time map because LOS effects due to the overlying plasmoid structures obscure it. The feature shifts downward along $y$ with increasing time, indicating the downward displacement of the nanojet-like signature as the upper flux rope proceeds downward towards the merging process. The brightening extends over 1–2 pixels along the $y$-axis, suggesting that the thickness of the feature varies between 600~km and 1.2~Mm. This suggests that the feature corresponds to a lifetime of $\approx 20$~s and a thickness of $\lesssim 1.2$~Mm. In contrast, this local brightening is absent in maps corresponding to hotter lines. The Fe XV (middle panel) and Fe XIX (right panel) maps highlight the upper and lower plasmoids from their pre-merger phase ($t \approx 90$–$120$~s) to the merged state ($t \gtrsim 135$~s). However, the intensity of the Fe XIX 108~\AA\ emission is about three orders of magnitude lower than that of the Fe XV 284~\AA\ line.      

The bright band seen below the ``nanojet-like signature" in the time-distance map (bottom-left panel of Figure~\ref{fig:synthetic_spectra}) arises from LOS integrated intensity across the entire horizontal domain (between $x=-20$ to $20$ Mm). Consequently, the appearance of the bright band might be a consequence due to the lower plasmoid (prior to merging), the merging plasmoid, the current sheet, and the background stratification, additionally with the substructures located away from $x=0$. Nevertheless, we consider such bright bands, either below or above the nanojet feature to be expected in observations of plasmoid-merging events. This interpretation is supported by the intensity maps in Figure~\ref{fig:nanojet_panels} (top row), particularly in the AIA 193 \AA\ and 211 \AA\ channels, where the plasmoid regions appear brighter than the nanojet location.

\subsection{Comparison with observation}\label{subsec:observation}

\begin{figure*}[hbt] 
    \centering
    \includegraphics[width=1\linewidth]{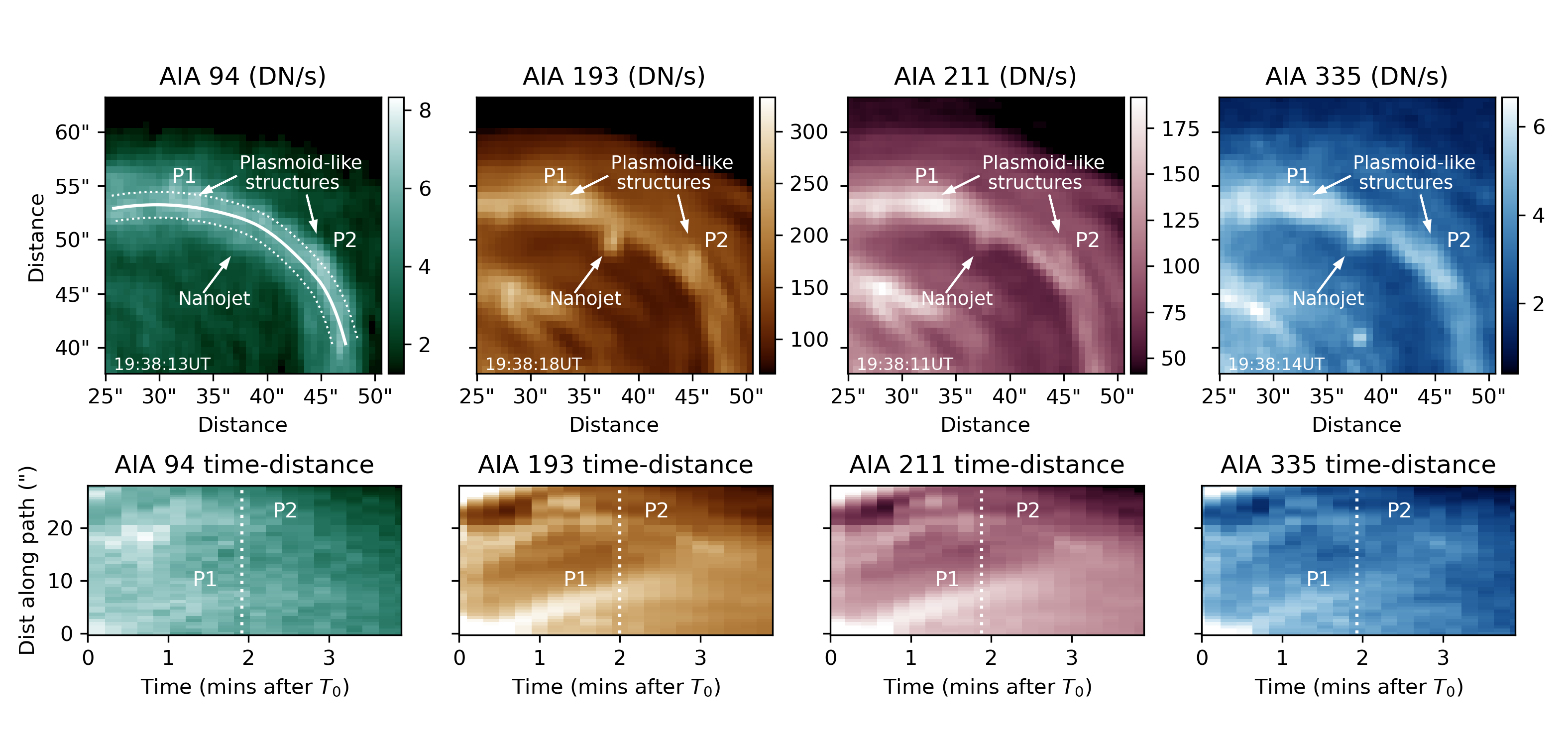}
    \caption{Top row: Selected AIA snapshots of nanojet N2 from \cite{Ramada:2022}, with arrows marking the plasmoid-like structures P1 and P2 in the strands moving towards each other. The white solid line is a path taken to produce the time-distance diagrams, where the value for a given point along the path is obtained by averaging eight values spaced evenly along its normal line between the dashed lines. The white vertical dashed line marks the timestamp of the images, where $T_0$ is at 19:36:18~UT. An animated version of this figure is available online.}
    \label{fig:nanojet_panels}
\end{figure*}

\begin{figure}[hbt] 
    \centering
    \includegraphics[width=1\linewidth]{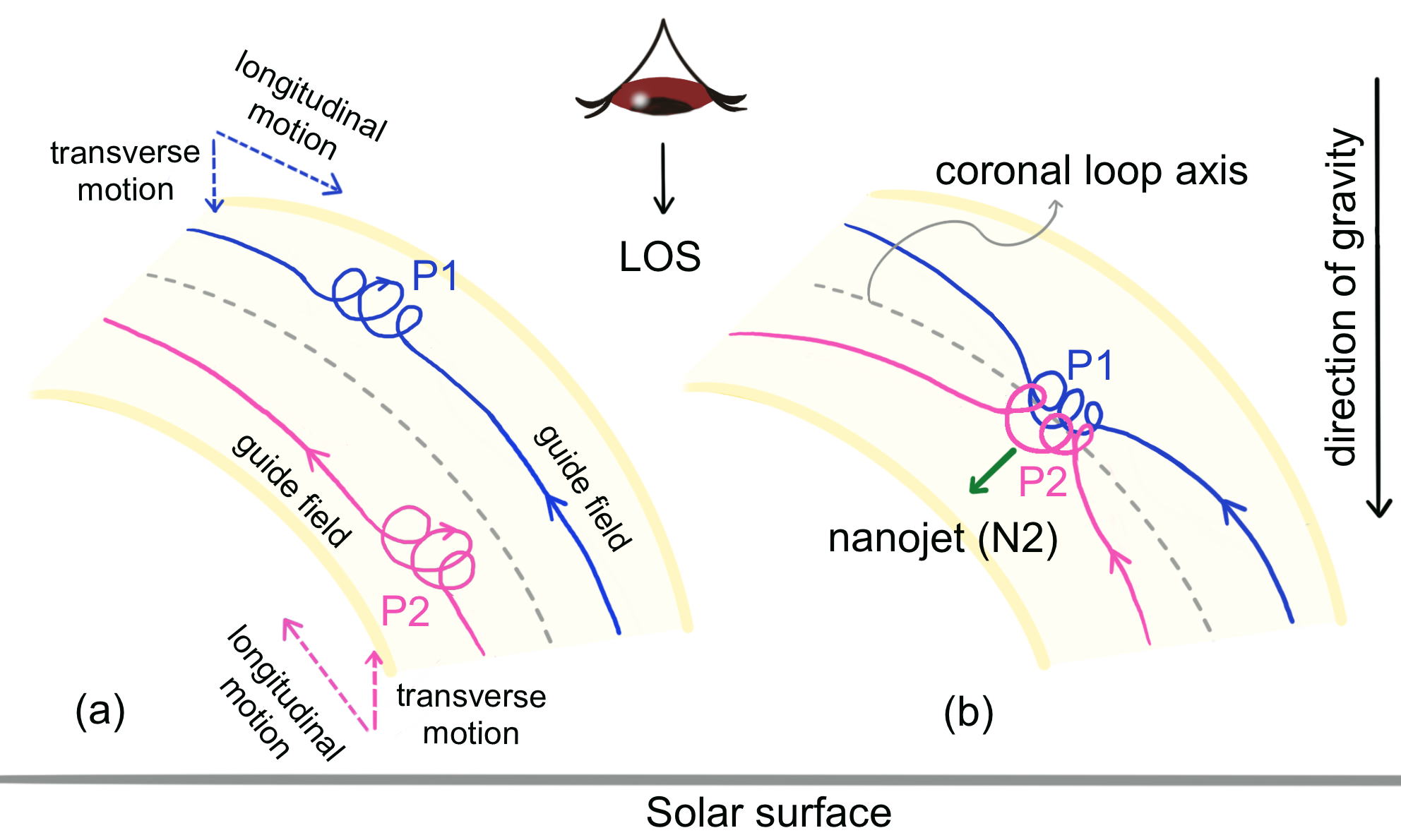}
    \caption{A schematic representation of the observation (top row of Figure~\ref{fig:nanojet_panels}) illustrating the plasmoid merging event and the subsequent nanojet (N2) ejection. The LOS of the observation is parallel to the direction of gravity. The yellow-shaded regions denote the coronal loop, with the two strands shown in blue and pink. Panel (a) depicts the pre-merger stage of plasmoids P1 and P2, which exhibit motion parallel to the guide-field direction (``longitudinal motion'') as well as (anti-)parallel to gravity (``transverse motion''). Panel (b) shows the interaction between plasmoids P1 and P2 at a later stage than shown in panel (a), leading to the ejection of the nanojet (N2).}
    \label{fig:cartoon}
\end{figure}

Nanojets have been observed in IRIS and AIA, with varying visibilities in the coronal channels of AIA \citep{Ramada:2022, Ramada:2024}. When large enough, they are always clearly visible in AIA 304~\AA\, and often also in 193~\AA\ and 171~\AA. For nanojets observed in hotter structures, we find that they can also be visible in the rest of the coronal AIA channels, but only faintly in AIA 94. An example from \cite{Ramada:2022} is shown in Figure \ref{fig:nanojet_panels}, of nanojet N2, which was observed at the outer edges of a loop-like structure powered by a blowout jet. N2 is 1.5~Mm in length, 1~Mm in width, with a lifetime of 20~s. Its temperature is estimated to be around 3.4~MK through the Differential Emission Measure analysis using the Basis Pursuit Method \citep{Cheung:2015}, and combined kinetic and thermal energy of $2.9\times10^{25}$~erg. In this case, the nanojet itself was found to be significantly brighter in IRIS and most of the hot AIA passbands, compared to its originating hot loop strands that are also visible in the majority of the hot AIA passbands. For this observation, we observed multiple misaligned strands (suggestive of braiding) with plasmoid-like structures in the blowout jet prior to the reconnection event. Figure \ref{fig:nanojet_panels} marks two such structures, where they are visible, specifically in the AIA 193, 211, and 335. These structures in the strands also appear to be moving towards each other before the nanojet occurrence, and then intercepting one another at the location of the nanojet (see the associated animation with Figure~\ref{fig:nanojet_panels}). 

A schematic illustration is presented in Figure~\ref{fig:cartoon}, depicting a plausible configuration of the observation (Figure~\ref{fig:nanojet_panels}) highlighting the motion of the plasmoid-like structures (P1 and P2), the associated guide fields, and the ejection of the nanojet (N2). The plasmoid-like structures are indicative of magnetic flux ropes containing guide fields oriented parallel to the coronal loop axis. This guide field also has a component perpendicular to the gravity, particularly at the locations of P1 and P2, as the observation was taken on-disk. These plasmoids can exhibit motion along the coronal loop axis (``longitudinal motion''), as well as parallel to the gravity (``transverse motion''), as shown in the panel (a) of Figure~\ref{fig:cartoon}. However, it is to be noted that the transverse motion of P1 and P2 cannot be inferred from the intensity maps (top row of Figure~\ref{fig:nanojet_panels}), because the LOS direction is (nearly) aligned with gravity. These plasmoids interact with each other at a later stage, leading to the ejection of nanojet (N2), as shown in panel (b) of Figure~\ref{fig:cartoon}. In comparison with the model by SMI25, the guide field acts solely in the direction perpendicular to the gravity. However, the bi-directional motion of the flux ropes in the model is (mainly) governed by the spatial inhomogeneity of the vertical component of the Lorentz force (see SMI25 for further details), indicating that the gravity plays a negligible role in driving the motion of the flux ropes.

The time-distance diagrams at the bottom rows in Figure~\ref{fig:nanojet_panels} show two slopes from the two plasmoid-like structures, indicating plasma motion that moves towards each other prior to the nanojet occurrence. We therefore suspect that the merging of the plasmoid-like structures, coupled with the misalignments between the strands, triggered the reconnection event which generated the nanojet. It is also possible that the merging plasmoids originated from reconnection between the Kelvin-Helmholtz instability (KHI) vortices, that is already present in the regions where the nanojets are observed \cite{Ramada:2022}.

In comparison with the synthetic observations produced from the model, we find that the observed nanojet (N2) and synthetic observations of nanojet-like features share similar morphology and intensity values in the AIA 193, 211, and 335 broadbands. We also notice that the length ($\lesssim$~2.4 Mm), width ($\lesssim 1.2$~Mm) and lifetime ($\approx 20$~s) obtained from the synthesis, and the temperature ($\lesssim 1.5$~MK) and energy flux ($\sim10^{24}$~erg) from the (MHD) model are in reasonable agreement with the observed nanojet.  

However, the observed nanojet (N2) ejection is unidirectional, propagating inward along the curvature radius of the hosting strands, probably facilitated by the dominant magnetic tension along the inward direction, whereas the modeled nanojet ejections are bi-directional, extending horizontally, with no directional bias due to the left–right symmetry of the configuration.

\section{Summary and Conclusion}\label{sec:conclusion}

In this Letter, we present, for the first time, the synthetic observations enabling spectroscopic diagnostics of nanojet-like features due to coalescence of two flux ropes (plasmoids in 2D) based on a 2.5D MHD model by SMI25. We perform forward modeling under the optically thin approximation to synthesize observables in multiple EUV channels corresponding to the current SDO/AIA, and the future MUSE missions. The synthetic emissivity maps in AIA 94, 193, 211, and 335~\AA\ passbands (bottom row of Fig.~\ref{fig:aia_synthesis}), and MUSE 171~\AA\ line (bottom-left panel of Fig.~\ref{fig:muse_synthesis}) reveal localized brightenings with lengths of $\lesssim 2.3$~Mm, reminiscent of nanojets. However, the AIA 335~\AA\, and MUSE 284, 108~\AA\ maps highlight the brightenings associated with the plasmoid structure. The spectral maps and line profiles in different emission lines (Fig.~\ref{fig:synthetic_spectra}) show bi-directional Doppler shifts of around $\pm 100$~km s$^{-1}$, reflecting the collimated flows along the direction of the nanojets ejection. The space-time intensity map (left panel of bottom row in Fig.~\ref{fig:synthetic_spectra}) indicate a lifetime of $\approx 20$~s, and thicknesses of $\lesssim 1.2$~Mm of the ejecting nanojets, whereas the space-time maps for other MUSE lines (middle and right panels of bottom row in Fig.~\ref{fig:synthetic_spectra}) show the signature of merging process of the plasmoids. The temperature and the total energy flux of the nanojets estimated from the MHD model are $\lesssim 1.5$~MK and $\approx 3.3\times 10^{24}$~erg. These estimated values from the synthesis and the simulation are in good agreement with the nanojet (N2) observation reported by \cite{Ramada:2022} in a blowout jet scenario, and also the morphological features of the nanojet and plasmoids (Fig.~\ref{fig:nanojet_panels}) show a reasonable similarity with the AIA synthetic maps (bottom row of Fig.~\ref{fig:aia_synthesis}).

In conclusion, this work underscores the importance of generating synthetic observables of nanojets to provide predictions for their detectability with current and future spectroscopic facilities. It also establishes an important connection between the modeling and observation of nanojets, opening avenues for a deeper understanding of the mechanisms that lead to nanojet formation, and for diagnosing their thermodynamics and energetics to assess their contribution to coronal heating.

\begin{acknowledgments}
We thank the anonymous referee for the useful and constructive comments that has improved the manuscript considerably. SS, DNS and FMI acknowledge support by the European Research Council through the Synergy Grant \#810218 (``The Whole Sun”, ERC-2018-SyG). They thankfully acknowledge the technical expertise and assistance provided by the Spanish Supercomputing Network (Red Espa\~{n}ola de Supercomputaci{\'o}n), as well as the computer resources used: the LaPalma Supercomputer, located at the Instituto de Astrof{\'i}sica de Canarias. JMS acknowledges support by NASA contract NNG09FA40C (IRIS) and 80GSFC21C0011 (MUSE) and NSF grants AGS2532363 and AGS2532187. MUSE is led by the Lockheed Martin Solar and Astrophysics Laboratory of Palo Alto, California. MUSE is managed by the Explorer’s Program Office of NASA’s Goddard Space Flight Center in Greenbelt, Maryland, for the Heliophysics Division of NASA’s Science Mission Directorate. Lockheed Martin Advanced Technology Center, along with partner institutions, builds the MUSE instrument and spacecraft and University of California, Berkeley provides the mission operations center. MUSE benefits from international contributions supported by the Norwegian Space Agency (NOSA), the Italian Space Agency (ASI), the German Space Agency at DLR, and from the Max Planck Institute for Solar System Research (MPS). The Solar Dynamics Observatory (SDO) is a mission for NASA's Living With a Star (LWS) program. AIA is an instrument on board the SDO. All SDO data used in this work are available from the Joint Science Operations Center (\texttt{http://jsoc.stanford.edu}) without restriction.

\end{acknowledgments}

\appendix 
\section{Forward modeling}\label{sec:forward modeling}
The emissivity ($\epsilon_{i,j}$) is estimated for each pixel labeled as the indices $i$ and $j$ assigned to the rows and columns, respectively,
\begin{align}
    \epsilon_{i,j} = A_b\ n_{e_{i,j}} n_{H_{i,j}} G(T_{i,j}),
\end{align}
where $A_b$ is the abundance of the emitting element, and $n_{e_{i,j}}$, $n_{H_{i,j}}$, $T_{i,j}$, and $G(T_{i,j})$ represent the electron and hydrogen number densities, temperature, and the contribution function at the $(i,j)$ cell, respectively. Here, we supply $n_{e_{i,j}}$, $n_{H_{i,j}}$, and $T_{i,j}$ obtained from the simulation. The contribution functions for the MUSE lines (Fe IX 171~\AA, Fe XV 284~\AA, and Fe XIX 108~\AA) are calculated for a temperature range between $10^5-10^8$ K with a constant plasma pressure of $3\times10^{15}$~K cm$^{-3}$. The contribution functions for the other lines and passbands are calculated for a temperature range between $10^5-10^8$ K with a constant density of $10^9$ cm$^{-3}$. The coronal elemental abundances are used from \cite{Feldman:92}. For single line emissivities, we use \texttt{ch\_synthetic.pro} from CHIANTI~10
\citep{DelZanna:2021}, while for AIA filters we use \texttt{aia\_get\_response.pro} from \texttt{SSWIDL} package. The temperature response of the AIA passbands is shown in \cite{Lemen_AIA:2012}, and for the MUSE lines in \cite{Cozzo:2024}. 
 
We calculate intensity summing over the rows for each column as,
\begin{align} \label{eq:spectral_response}
  I_j = \sum_i  \phi_f(\lambda) \epsilon_{i,j}\ \Delta l_{i,j},  
\end{align}
where, $\Delta l_{i,j}$ is the length of the corresponding cell along the LOS direction (along the rows in this case), and $\phi_f(\lambda)$ is spectral response function of the $f$-th instrument channel (similar to Equation 3 of \cite{Zacharias:2011}, which was formulated in the velocity space, instead),
\begin{align}
    \phi_f(\lambda) = \frac{1}{\sqrt{\pi}\ \Delta\lambda} \exp\bigg[-\bigg(\frac{\lambda - \lambda_s}{\Delta \lambda}\bigg)^2\bigg],
\end{align}
where, 
\begin{align}
    \lambda_s = \lambda_0 (1+v_{LOS_{i,j}}/c)
\end{align}
is the shifted wavelength from the central wavelength ($\lambda_0$) of the emission line, $v_{LOS_{i,j}}$ is the LOS velocity at the $(i,j)$ pixel, $c$ is the velocity of light in vacuum, and 
\begin{align}\label{eq:thermal_broadening}
    \Delta \lambda = \frac{\lambda_0}{c}\sqrt{\frac{2 k_B T_{i,j}}{m}},
\end{align}
is the wavelength shift due to thermal broadening, 
\begin{align}\label{eq:vth}
    v_{th} = \sqrt{\frac{2 k_B T_{i,j}}{m}},
\end{align}
where $k_B$ is the Boltzmann constant, and $m$ is the atomic mass of the emitting element. The intensity of the emitting plasma at the $(i,j)$ cell is
\begin{align}
    F_{i, j} = A_b\ n_{e_{i,j}} n_{H_{i,j}} \Lambda_f(T) \Delta l_{i,j},
\end{align}
where, $\Lambda_f(T)$ is the temperature response function per pixel, which combines the emission properties of the plasma with the response of the $f$-th instrument channel, $\phi_f(\lambda)$,
\begin{align}
    \Lambda_f(T) = A_{pix}\int_\lambda G_\lambda(T) \phi_f(\lambda) \rm{d}\lambda,
\end{align}
where $A_{pix}$ is the area of a single cell. The total line width due to thermal, non-thermal, and instrumental broadening can be calculated through the second moment of the LOS velocity, which is estimated for each column as,
\begin{align}\label{eq:2nd_mom}
    \Delta\bar{v}_j = \sqrt{\frac{\sum_i\ (v_{i,j} - \bar{v}_j)^2 F_{i,j}}{\sum_i F_{i,j}}},
\end{align}
where, $\bar{v}_j$ is the first moment of the velocity distribution defined as
\begin{align}
    \bar{v}_j = \frac{\sum_i v_{i,j} F_{i,j}}{\sum_i F_{i,j}}. 
\end{align}

\bibliography{reference}{}
\bibliographystyle{aasjournalv7}

\end{document}